\documentclass[onecolumn]{aa}
\usepackage{amsmath,amssymb}
\usepackage{hyperref}
\usepackage{graphicx}
\usepackage{url}

\def\apjl{Astrophys. J. Lett.}
\def\apjs{Astrophys. J. Suppl.}
\def\ao{Appl. Opt.}

\def\aap{Astron. Astrophys.}

\newcommand{\Cont}{\mathcal{F}}
\newcommand{\Stream}{\mathcal{M}}
\newcommand{\PA}{\mathcal{G}}

\begin{document}
\title{Stochastic acceleration in arbitrary astrophysical environments}
\author{Dominik Walter \inst{\ref{inst1}, \ref{inst2}} \& Björn Eichmann \inst{\ref{inst1}, \ref{inst2}}}
\institute{Institut f\"ur Theoretische Physik IV, Ruhr-Universit\"at Bochum, 
Universit\"atsstrasse 150, 44780 Bochum, Germany \label{inst1} \and Ruhr Astroparticle and Plasma Physics Center (RAPP Center), 44780 Bochum, Germany\\ \email{dw@tp4.rub.de; eiche@tp4.rub.de } \label{inst2} }

\abstract{Turbulent magnetic fields are to some extent a universal feature in astrophysical phenomena. Charged particles that encounter these turbulence get on average accelerated according to the so-called second-order Fermi process. However, in most astrophysical environments there are additional competing processes, such as different kinds of first-order energy changes and particle escape, that effect the resulting momentum distribution of the particles. 

In this work we provide to our knowledge the first semi-analytical solution of the isotropic steady-state momentum diffusion equation including continuous and catastrophic momentum changes that can be applied to any arbitrary astrophysical system of interest. Here, we adopt that the assigned magnetic turbulence is constrained on a finite range and the particle flux vanishes beyond these boundaries. Consequently, we show that the so-called pile-up bump---that has for some special cases long been established---is a universal feature of stochastic acceleration that emerges around the momentum $\chi_{\rm eq}$ where acceleration and continuous loss are in equilibrium if the particle's residence time in the system is sufficient at $\chi_{\rm eq}$. 
In general, the impact of continuous and catastrophic momentum changes plays a crucial role in the shape of the steady-state momentum distribution of the accelerated particles, where simplified unbroken power-law approximations are often not adequate. }
\keywords{stochastic acceleration -- high energies -- turbulence } 
\maketitle

\section{Introduction}\label{Sec:Intro}
Stochastic acceleration of charged energetic particles has been the initial explanation \citep{Fermi1949} for the formation of power-law structure in the detected cosmic-ray (CR) energy spectra. Since its momentum change $\delta p/p\propto (v_{\rm A}/c)^2$ is quadratic with respect to the characteristic velocity of the ``accelerator" (which is e.g.\ the Alfv\'en velocity of the magnetic turbulence), this is typically called ``second-order Fermi" or short ``Fermi-II" process. However, for non-relativistic turbulence and weakly magnetized plasma the acceleration by a shock, with velocity $v_{\rm sh}$, is typically more efficient yielding $\delta p/p\propto v_{\rm sh}/c$, which is why this is referred as the ``first-order Fermi" or short ``Fermi-I" process. In both cases the characteristic length scale of the ``accelerator" needs to be on similar scales as the gyro-radius of the particle that shall be addressed, i.e.\ the particle's scattering mean free path needs to be within the resonant scattering regime. Due to the so-called turbulent cascade \citep{Kolmogorov1941, Kraichnan1965} the spectrum of magnetic turbulence typically spans over several orders of magnitude before the turbulence loses its energy through dissipation, so that Fermi-II naturally works for a broad range of particle energies. So there are numerous astrophysical environments, such as accretion disks \citep{Liu2006}, solar flares \citep{Petrosian1999}, AGN coronae \citep{Fiorillo+2024, Murase+2020} or blazars \citep{Kataruski} where it has been shown that the acceleration of CRs is effected by the Fermi-II process. In particular the recent observation of high-energy neutrinos from the Seyfert galaxy NGC\,1068 by the IceCube experiment \cite{IceCube2022_ngc1068}, that most likely originate in the corona of its AGN \citep[e.g.][]{Eichmann+2022}, highlights the importance of a proper description of the CR momentum distribution---within environments where competing gain and loss processes are at work---to understand its resulting messengers. 
However, to accelerate CRs from thermal energies up to hundreds of TeV or more, the associated range of its scattering mean free path indicates that in many of these astrophysical applications multiple acceleration scenarios can be involved. Moreover, there are also other phenomena such as reconnection-driven particle acceleration \citep[e.g.][]{XuLazarian2023,ComissoSironi2018,LazarianVishniac1999} or magnetic trapping \citep[e.g.][]{Pezzi+2022,Trotta+2020} that can mimic the Fermi-I process without the presence of an actual shock structure. Hence, the overall acceleration of CRs up to hundreds of TeV and more involves most likely a combination of different processes whose relevance dependent on the particle's momentum $p$. 
A well-established approach to determine the evolution of the CR particle in phase space is to reduce the Fokker-Planck equation for the case of an ensemble-averaged particle distribution $\langle f \rangle$, which is supposed to be spatially uniform and isotropic in $p$, hence $\langle f(\vec{r},\,\vec{p},\,t) \rangle = \langle f(p,\,t) \rangle$, so that we obtain the so-called momentum diffusion equation
\begin{equation}
{\partial \over \partial t} \langle f(p,\,t) \rangle = {1 \over p^2} \, {\partial
\over \partial p} \, \left[ p^2 \, D(p) \, {\partial \over \partial p} \,
\langle f(p,\,t) \rangle \right] \, .
\label{monDiffEq}
\end{equation}
Here, $D(p)\propto p^q$ denotes the momentum diffusion coefficient which approximates the interaction rate with the turbulence. But for most astrophysical systems, as we have just discussed, this description is not sufficient and we need to incorporate also the impact of continuous losses and additional gains (e.g. by Fermi-I processes) of momentum, hereafter referred to as $\langle \dot{p} \rangle$, as well as catastrophic losses on a timescale $t_{\rm esc}(p)$ such as by the escape from the physical system. Note that these coefficients are in general also time-dependent, but if the timescale of interest $t$ is significantly shorter than the associated variations of the physical system, we can assume that these coefficients are independent of the time.
Injecting particles according to the source term $\tilde{Q}(p,\,t)$ into the system, we thus obtain that the isotropic particle momentum distribution $n(p,\,t)=4\pi p^2\,\langle f(p,\,t) \rangle$ is given by
\begin{equation}
    \frac{\partial n}{\partial t} =\frac{\partial }{\partial p} \left(  D \left( p \right) \frac{\partial n}{\partial p} \right) -\frac{\partial }{\partial p} \left( \left( \frac{  2 D \left( p \right)}{p} + \langle \dot{p} \rangle \right) n \right) -\frac{n}{t_{\rm esc}(p)} + \tilde{Q}\,.
    \label{OrigTranspEq}
\end{equation}
First this equation has been solved by \cite{Davis1956} and \cite{Achterberg1979} for the special case of vanishing additional additional energy changes ($\langle \dot{p} \rangle=0$), diffusive escape and the so-called ``hard-sphere'' approximation ($q=2$), where the mean free path for the particle-wave interaction becomes independent of the particle momentum. This leads at a momentum $p$ larger than the injected momentum $p_{\rm inj}$ to $n(p>p_{\rm inj})\propto p^{\sigma}$, with $\sigma=0.5-\sqrt{(9/4)+\varepsilon}$. Hence, the resulting particle momentum distribution depends on the parameter $\varepsilon = \tau _{\text{acc}}/\tau _{\text{esc}}$, which denotes the ratio of acceleration and escape timescales at some chosen particle momentum $p_0$. 
Neglecting also the particle escape, it has been shown by e.g.\ \cite{Lacombe1979} or \cite{DroegeSchlickeiser1986} that the isotropic particle momentum distribution yields $n(p>p_{\rm inj})\propto p^{1-q}$ for momentum dependent particle-wave interaction in the range $1\leq q < 2$. 
Consequently, these works already showed that the ``typical'' particle momentum distribution from stochastic acceleration is significantly harder than what is expected from a ``typical'' Fermi-I process \citep[e.g.][]{Drury1983, Ellison1990}. 
Moreover the analytical solutions for various special cases of the continuous momentum losses have been investigated, whereof one of the most generalised efforts have been made by \cite{ParkPetrosian1995} (that also provide a very detailed historical review of the different approaches in the previous century to solve Eq.~\ref{OrigTranspEq}) and \cite{StawPetro2008} (hereafter SP08). One of the striking differences between these two investigations is the adopted boundary condition as well as the momentum range of the particle: In \cite{ParkPetrosian1995} the ``no-flux'' boundary condition is applied for an infinite range of momentum ($0\leq p \leq \infty$), hence $\Cont(p=0,t)=\Cont(p=\infty,t)=0$, leading to a singular problem, where steady state solutions only exist under certain conditions. In contrast, SP08 restrict their analysis to a finite range ($p_1\leq p \leq p_2$) since the adopted momentum diffusion coefficient $D(p)$ is only valid in a limited range where particle-wave interactions take place. In doing so, the ``no-flux'' boundary condition is only applied for certain cases that account for a non-vanishing particle escape ($\varepsilon \neq 0 $). But in theses cases analytical solutions are determined only in the ``hard-sphere'' approximation for certain momentum loss processes of ultrarelativistic electrons.

In this paper we attempt to give solutions for the most general case of the steady state form of Eq.~\ref{OrigTranspEq} and discuss a number of different cases for the loss processes. To achieve this we first introduce in Sect.~\ref{Sec:AnalyticalApproach} the original approach given by SP08 and the used modifications. In Sect.~\ref{Sec:Semi-AnaSol} we determine a semi-analytical solution, where the problem has been simplified to a so-called Riccati equation that can be tackled easily by a numerical solver. Moreover, we provide in Sect.~\ref{Sec:AnaSolRiccati} an analytical treatment of this remaining equation in a limited range of parameters. In Sect.~\ref{Sec:Results} we provide the resulting momentum distribution for some dedicated scenarios and compare our results with previous investigations. Finally, we summarize and discuss our findings in Sect.~\ref{Sec:Summary}. \\

\section{Analytical approach}\label{Sec:AnalyticalApproach}
In the following we assign the presence of an isotropic Alv\'enic turbulence with a wave energy spectrum of the form $\mathcal{W}  \left( k \right) \propto k^{-q}$ in a finite range of wavenumbers ($k_1\leq k \leq k_2$). Moreover, we consider the spectral index in the range $1\leq q \leq 2$ and suppose that the energy density in the turbulent magnetic field $\delta B$ is small compared to the regular magnetic field $B_0$, so that $\zeta \equiv (\delta B)^2/B_0^2<1$. In that case, the momentum diffusion coefficient yields \citep[e.g.][]{Melrose1968, Schlickeiser1989}
\begin{equation}
    D(p)\simeq \frac{\zeta \beta_{\rm A} p^2 c}{r_{\rm g}^{2-q} (2\pi/k_1)^{q-1}}\propto p^q\,,
\end{equation}
where $\beta_{\rm A}=v_{\rm A}/c$ denotes the dimensionless Alvf\'en velocity and $r_{\rm g}=pc/(eB_0)$ is the gyro-radius of the relativistic particles of interest. The characteristic acceleration time due to stochastic particle-wave interactions is given by $t_{\rm acc}=p^2/D(p)$. 
To facilitate the comparison with the work by SP08, we also introducing the dimensionless momentum variable $\chi\equiv p/p_0$, where $p_0$ is chosen to be the injection particle momentum. First-order momentum changes according to a characteristic continuous energy loss/gain-timescale $t_{\rm loss/gain}(\chi)$ are hereby given by $\vartheta _{\chi}\equiv-\langle \dot{p} \rangle\,\tau_{\rm acc}/p = \pm \tau_{\rm acc}/t_{\rm loss/gain}(\chi)$, where 
\begin{equation}
    \tau_{\rm acc} \equiv \frac{2\pi}{\zeta\beta_{\rm A}^2c k_1}\,\left( \frac{p_0c k_1}{2\pi eB_0} \right)^{2-q} \, ,
\end{equation}
so that $t_{\rm acc}(\chi)=\tau_{\rm acc}\,\chi^{2-q}$. Moreover, we define the dimensionless steady state particle momentum distribution $N(\chi)\equiv p_0\, n(p)\, V$ and the corresponding continuous source $Q(\chi)\equiv \tau_{\rm acc}\,p_0\, \tilde{Q}(p)\, V$, where $V$ denotes the volume of the system of interest, so that the steady state momentum diffusion Eq.~\ref{OrigTranspEq} becomes
\begin{equation}
    \frac{\partial }{\partial \chi} \left( \chi ^q \frac{\partial N}{\partial \chi}\right) - \frac{\partial }{\partial \chi} \left[  \left( 2\chi ^{q-1} - \chi \vartheta _{\chi} \right) N\right] - \varepsilon_\chi\,N = -Q \, .
    \label{Eq:BaseEquation}
\end{equation}
Here, $\varepsilon_\chi\equiv\tau_{\rm acc}/t_{\rm esc}(\chi)$ denotes the catastrophic loss rate, which in case of diffusive escape from the system yields (see e.g.\ SP08) $\varepsilon_\chi=\tau_{\rm acc} \chi^{2-q}/\tau_{\rm esc}$, where 
\begin{equation}
    \tau_{\rm esc} \equiv \frac{9L^2\,\zeta\,k_1}{2\pi\,c}\,\left( \frac{p_0c k_1}{2\pi eB_0} \right)^{q-2}\, .
\end{equation}
For the case of $\varepsilon_\chi = 0$ the general solution for any arbitrary form of $\vartheta _x$ is given in SP08. We now aim to determine a solution for the case $\varepsilon_\chi \neq 0$. For this purpose we will use the results given in that work as a base and modify it for non-vanishing $\varepsilon_\chi$-values.
\\
We start in a similar manner as SP08 by transforming the differential equation into a more suitable form by introducing the term 
\begin{equation}
    S\left( \chi \right) = \exp \left\lbrace  -\int_{\chi _1}^{\chi} 2 \chi '^{-1} - \chi '^{1-q} d\chi '  \right\rbrace \, ,
\end{equation} 
so that Eq.~\ref{Eq:BaseEquation} reads as 
\begin{equation}
    \frac{\partial }{\partial \chi} \left( S \left( \chi \right) \chi ^{q} \frac{\partial N}{\partial \chi} \right) - \PA \left( \chi \right)  N = -Q\left( \chi \right) S \left( \chi \right) \,,
    \label{Eq:Main-Equation}
\end{equation}
with
\begin{equation}
    \PA \left( \chi \right) = \left( 2 \left( q-1 \right) \chi ^{q-1}- \frac{\partial }{\partial \chi} \left( \chi \vartheta _{\chi } \right) + \varepsilon_\chi \right) S\left( \chi \right) \,.
\end{equation}
For $\PA \left( \epsilon_\chi = 0 \right)$ SP08 provide the solutions 
\begin{align}
    y_{1,\varepsilon_\chi = 0} &= S^{-1} \left( \chi \right) \\
    y_{2,\varepsilon_\chi = 0} &= S^{-1} \left( \chi \right) \int_{\chi_1}^{\chi} \chi'^{-q} S\left(  \chi' \right) d\chi'
\end{align}
of the homogeneous equation, which can be used to construct a suitable Green's function. 

For the case $\varepsilon_\chi \neq 0$, we adopt a slightly modified function to include the additional dependency.
Hence, we use a modification according to
\begin{equation}
    \tilde{S } \left( \chi \right) = S \left( \chi \right)  \exp \left\lbrace \int _{\chi _1}^{\chi} m \left( \chi ' \right) d\chi ' \right\rbrace \label{Smod}  \, ,  
\end{equation}
so that the first solution is still given by $y_1 = \tilde{S}^{-1} \left( \chi \right)$.
To make certain calculations further down the line a bit simpler and without limiting the general character of the solution, we can assume, that $m \left( \chi _1 \right) = 0$, a brief discussion of the case $m \left( \chi _1 \right) \neq 0$ will be at the end of  Sec.~\ref{App:GenralStreaming}.
If we insert our form of $y_1$ into the differential operator $\frac{\partial }{\partial \chi} \left( S \left( \chi \right) \chi ^{q}  \frac{\partial }{\partial \chi} \right)$, we derive
\begin{align}
      \frac{\partial }{\partial \chi} \left( S \left( \chi \right) \chi ^{q} \frac{\partial \tilde{S} ^{-1} }{\partial \chi} \right) &= -m\left( \chi \right) \left( 2 \chi ^{q-1} - \chi \vartheta _{\chi}  \right) \exp \left\lbrace - \int _{\chi _1}^{\chi} m \left( \chi ' \right) d\chi '\right\rbrace \notag \\
      &+ \left( 2 \left( q-1 \right)  \chi ^{q-2} \frac{\partial }{\partial \chi} \left( \chi \vartheta _{\chi} \right)\right) \exp \left\lbrace - \int _{\chi _1}^{\chi} m \left( \chi ' \right) d\chi '\right\rbrace \notag \\
      &- \frac{\partial }{\partial \chi} \left( m \left( \chi \right) \chi ^q  \exp \left\lbrace - \int _{\chi _1}^{\chi} m \left( \chi ' \right) d\chi '\right\rbrace \right) \\
      &= -m\left( \chi \right) \left( 2 \chi ^{q-1} - \chi \vartheta _{\chi}  \right) \exp \left\lbrace - \int _{\chi _1}^{\chi} m \left( \chi ' \right) d\chi '\right\rbrace \notag \\
      &+ \left( \PA \left( \chi \right) S^{-1}\left( \chi \right) - \varepsilon _{\chi} \right) \exp \left\lbrace - \int _{\chi _1}^{\chi} m \left( \chi ' \right) d\chi '\right\rbrace \notag \\
      &- \frac{\partial }{\partial \chi} \left( m \left( \chi \right) \chi ^q  \exp \left\lbrace - \int _{\chi _1}^{\chi} m \left( \chi ' \right) d\chi '\right\rbrace \right) \\
      &  \overset{!}{=} \PA \left( \chi \right) \tilde{S} ^{-1} \left( \chi \right) \, .
\end{align}
So it follows that $y_1 = \tilde{S}^{-1}\left( \chi \right)$ solves the homogeneous form of Eq.~\ref{Eq:BaseEquation} if the modification function $m(\chi)$ obeys the following Riccati equation 
\begin{equation}
    \frac{d m}{d \chi }= m^2 -m \left( 2 \chi ^{-1} - \chi ^{1-q} \vartheta _{\chi} + q \chi ^{-1}  \right)  - \varepsilon _{\chi} \chi ^{-q} \,.
    \label{Eq:Riccati}
\end{equation}
This yields a tremendous simplification of the original problem and there are different strategies on how to solve this first-order ordinary differential equation as illustrated in Sect.~\ref{Sec:Semi-AnaSol}.

But first, a second solution $y_2$ of the homogeneous problem (\ref{Eq:BaseEquation}) needs to be determined. For this purpose we make a very similar attempt and modify the solution $y_{2,\varepsilon_\chi = 0}$ from SP08 as follows:
\begin{equation}
   y_2 = \tilde{S}^{-1}\left( \chi \right)\int\limits_{\chi _1}^{\chi } \chi '^{-q}\tilde{S} \left( \chi ' \right) \exp \left\lbrace \int\limits_{\chi _1}^{\chi '} m\left( \chi '' \right) d\chi ''  \right\rbrace  d\chi ' \, .
\end{equation}
This form of the solution can again be obtained by multiplying another modification term to the $y_{2,\varepsilon_\chi = 0}$ solution given in SP08 and solving the resulting equation for this modification.  
After inserting this ansatz for $y_2$ into the homogeneous form of Eq.~\ref{Eq:Main-Equation} one gets
\begin{align}
    \frac{\partial }{\partial \chi } \left(  \chi ^q S \left( \chi \right) \frac{\partial y_2}{\partial \chi }\right) = &\:\frac{\partial }{\partial \chi} \left( S \left( \chi \right) \chi ^q \frac{\partial \tilde{S}^{-1}\left( \chi \right)}{\partial \chi} \right) \int\limits_{\chi _1}^{\chi} \chi '^{-q}\,S \left( \chi ' \right)\, M_2\left( \chi ' \right) d\chi '  \notag \\
    &+ \tilde{S}^2\left( \chi \right) \frac{\partial \tilde{S}^{-1}\left( \chi \right)}{\partial \chi} + \frac{\partial \tilde{S} \left( \chi \right)}{\partial \chi}  \notag \\
    &= \PA \left( \chi \right)  y_2 \, .
\end{align}

To determine all solutions to the homogeneous form of Eq.~\ref{Eq:BaseEquation} a linear combination of these solutions $y_1$ and $y_2$ can be used. With the help of those solutions  one can derive the problems Green's function $G$ which can subsequently be used to determine the steady state particle momentum distribution $N(\chi)$ as shown in the following.

\section{Semi-analytical solution}\label{Sec:Semi-AnaSol}
Now that we have found two linear independent solutions of the homogeneous equation of transport (\ref{Eq:BaseEquation}) we can combine them into a Green's function $G$ by the approach
\begin{align}
    \frac{\partial }{\partial \chi} \left( S \left( \chi \right) \chi ^{q} \frac{\partial G}{\partial \chi} \right) - \PA G &= \delta \left( \chi - \chi _0 \right) \\
    G &= C\left( \chi \right)u_1 + D\left( \chi  \right) u_2 \\
    u_1 &= y_1 + \alpha y_2 \\
    u_2 &= y_1 + \beta y_2
\end{align}
in an analogous way to SP08. Therefore we also get an analogous result for the Green's function itself
\begin{equation}
    G\left( \chi , \chi_0 \right) = \frac{1}{w\left( \chi_0 \right) S \left( \chi_0 \right) \chi_0^q} \left[ u_1\left(  \chi \right) u_2\left( \chi_0 \right) H \left(  \chi_0 - \chi \right) + u_1\left(  \chi_0 \right) u_2\left( \chi \right) H \left(  \chi - \chi_0 \right) \right] \, ,
    \label{Eq:Green0}
\end{equation}
where we used the Wronskian determinant of $u_1$ and $u_2$, so $w\left( \chi \right) = u_1 u'_2 - u'_1 u_2$. All solutions to Eq.(\ref{Eq:Main-Equation}) are now given by
\begin{equation}
    N \left( \chi \right) = -\int\limits_{\chi _1}^{\chi _2 } G \left( \chi , \chi_0 \right)   S\left( \chi_0 \right) Q \left( \chi_0 \right) d\chi_0 \, . \label{Eq:GenSol}
\end{equation}
The only things left to determine are the values of the parameters $\alpha$ and $\beta$. As in SP08 we now use the opportunity to determine these two parameters by studying the continuity equation, that arises when we integrate Eq.~\ref{Eq:BaseEquation} over our entire domain, from $\chi _1$ to $\chi _2$. It follows for the stationary equation that
\begin{equation}
    \Cont \left( N \right) \mid _{\chi _2} - \Cont \left( N \right) \mid _{\chi _1} = \int\limits_{\chi _1 }^{\chi _2} \left( Q \left( \chi _0 \right) - \varepsilon _{\chi _0 } \right)  d\chi _0 \,  ,\label{Eq:StreamProblem}
\end{equation}
whereby the streaming operator $\Cont$ is given by
\begin{align}
    \Cont &=  \left( 2 \chi ^{q-1} - \chi \vartheta _{\chi} \right)  - \chi ^q \frac{\partial    }{\partial \chi} \,.\label{Op:Streaming}\\
\end{align}
All details of the evaluation of this boundary condition can be found in the Appendix. It leads to the two solutions 
\begin{align}
    \alpha  &= 0 \\
    \beta &= \frac{m \left( \chi _2 \right)\chi _2 ^q  y_1 \left( \chi _2 \right)}{\exp \left\lbrace \int\limits _{\chi _1}^{\chi _2} m \left( \chi _0 \right) d\chi _0 \right\rbrace - m \left( \chi _2 \right) \chi _2 ^q  y_2 \left( \chi _2 \right)}\,, \label{Eq:StramSolution}
\end{align}
that can be used to complete the Green's function. 
Hence, the Green's function (\ref{Eq:Green0}) of the problem is finally given by
\begin{equation}
    G(\chi,\,\chi_0) = - \frac{\exp\left[ -\int_{\chi_1}^{\chi_0} m(\chi')\,d\chi' \right]}{\beta\,\tilde{S}(\chi)}\, \Bigg\{ 1 + \beta \int_{\chi_1}^{\text{max}(\chi,\chi_0)}\tilde{S}(\chi')\,\chi'^{-q}\, \exp\left[ \int_{\chi_1}^{\chi'} m(\chi'')\,d\chi'' \right]\,d\chi' \Bigg\} \,.
    \label{Eq:Green_final}
\end{equation}
Apart from an analytical solution for $m(\chi)$, which will be discussed in the following section, we have determined all ingredients to calculate the steady state particle momentum distribution (\ref{Eq:GenSol}) for any arbitrary physical condition. Note that a few of these expressions tend to become huge and tiny numbers, respectively, that need to be summed up properly by the numerical algorithm.

\subsection{Analytical solution of $m(\chi)$}\label{Sec:AnaSolRiccati}
\begin{figure}[htb]
\centering
    \includegraphics[width=0.26\textwidth]{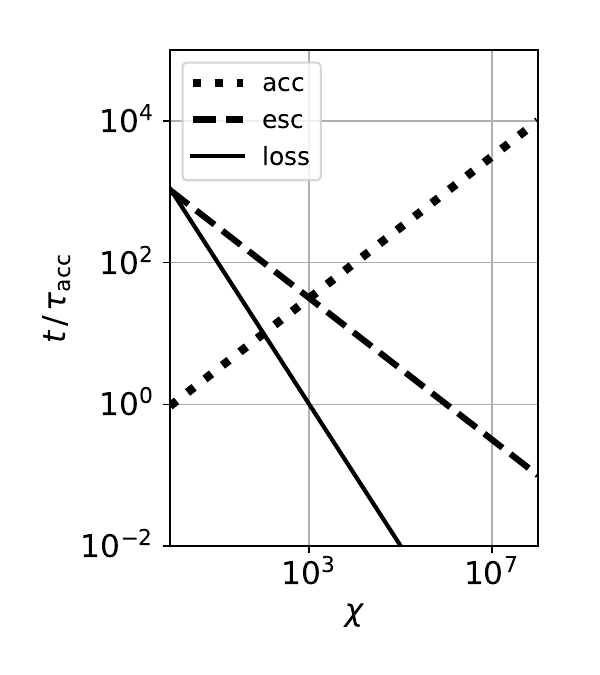}
    \includegraphics[width=0.45\textwidth]{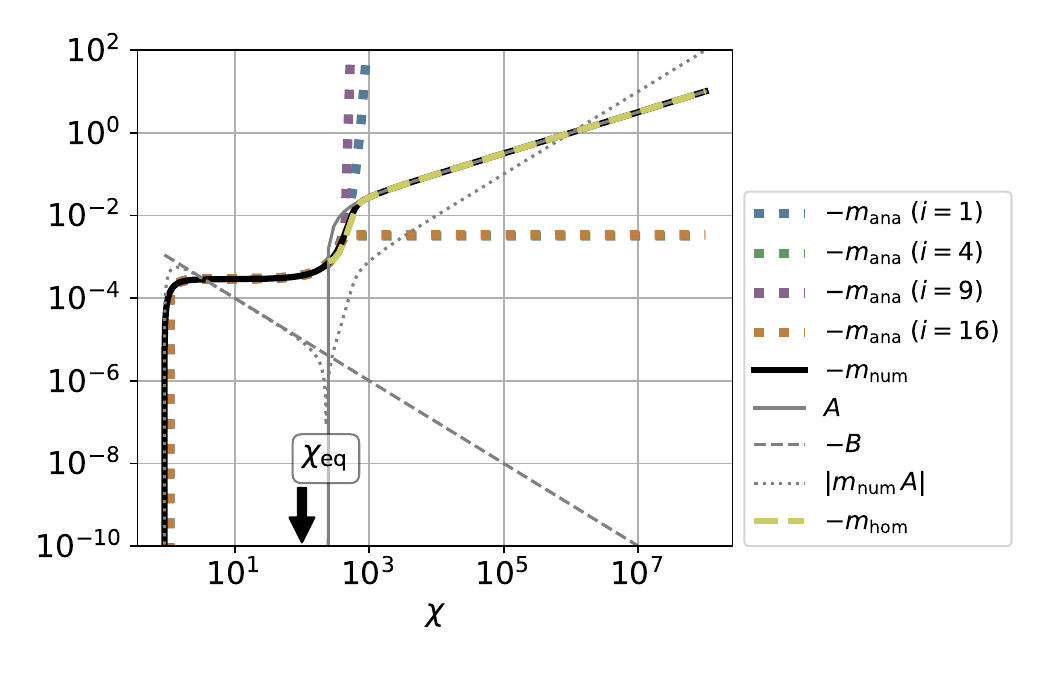}
    \hspace{-.3cm}\includegraphics[width=0.275\textwidth]{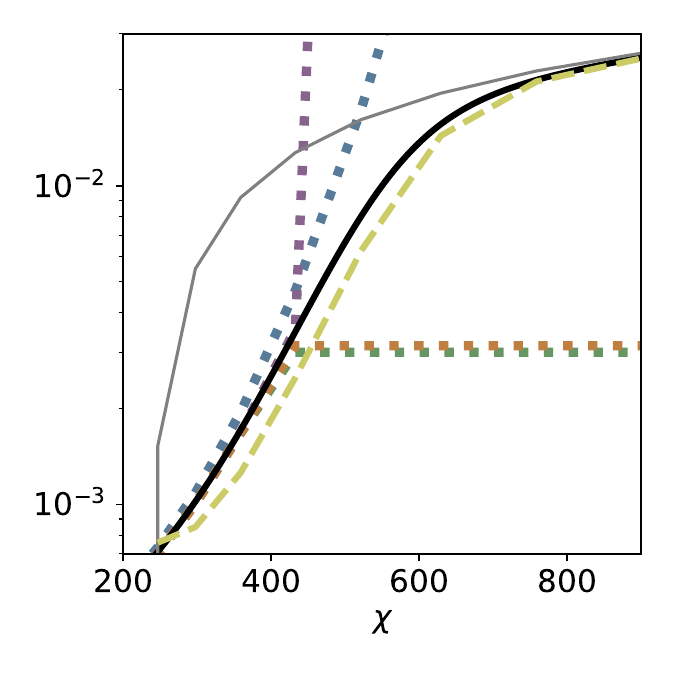}
        \caption{The characteristic timescales (left plots) for an exemplary scenario and the resulting modification function $m$ (middle and right plots) using either a numerical solver (black solid line) or an analytical approximation (colored lines). Further the spectral behavior of $A$ and $B$ is shown (grey lines). Here we adopted that $\chi _{\ast}$ satisfies $A\left( \chi _{\ast} \right) = 0$.}
        \label{Fig:m_ana}
\end{figure}

We will call 
$A = \chi ^{1-q} \vartheta _{\chi} -  (2+q) \chi ^{-1} $
and $B = - \varepsilon _{\chi} \chi ^{-q}$ and write the differential equation (\ref{Eq:Riccati}) for $m$ as follows 
\begin{equation}
    \frac{d m   }{d \chi} = m^2 + A \left( \chi \right) m + B \left( \chi \right) \label{Eq:mequation} \, .
\end{equation}
This resembles in the most general case a Riccati differential equation. 
The fact that we established $m\left( \chi _1  \right) = 0$ earlier in the text, provides us with a unique solution for the function $m$. The solution to Eq.(\ref{Eq:mequation}) can be found by very basic numerical procedures\footnote{We used the \emph{solve\_ivp} function, which is provided by the \href{https://docs.scipy.org/doc/scipy/reference/generated/scipy.integrate.solve_ivp.html}{\emph{scipy.integrate}} package.}. It is, however, also possible to derive approximations for very general cases, which will be discussed over the next paragraphs. Let us first consider the general case of $\varepsilon_{\chi } = \varepsilon _{\chi}\left( \chi\right)$, and assume that it represents a small perturbation. 
 If $\varepsilon _{\chi} \neq 0$ then there exists a value $\hat{\chi } $  so that $\varepsilon _{\chi} \left( \hat{\chi } \right) = \hat{\varepsilon } \neq 0$. 
 We can use the value $\hat{\varepsilon }$ as a small constant, in which we expand $m$ as a series, hence
 \begin{equation}
     m = \sum\limits _{i=0}^{\infty} \hat{\varepsilon } ^i m_i\, , \quad \text{so that }\quad 
     m^2 = \sum\limits_{i=0}^{\infty} \hat{\varepsilon} ^i \sum\limits_{k=0}^{i} m_im_{k-i}\, .
 \end{equation}
 Since $\hat{\varepsilon }$ is constant it will not be affected by differentiation and we, therefore, end up with a number of differential equation, where we use the function $b\left( \chi \right) = \hat{\varepsilon}^{-1}B \left( \chi \right)$ and get
\begin{align}
    \frac{dm_0}{d\chi} &= m_0^2 -A\left( \chi \right) m_0\, , \\
    \frac{dm_1}{d\chi} &= 2m_0m_1 - A\left( \chi \right) m_1 - b\left( \chi \right)\, , \\
    \frac{d m_i}{d \chi} &= \left( \sum\limits_{k=0}^i m_km_{i-k} \right)  -A\left( \chi \right) m_i \, .
\end{align}
Due to the fact, that $m \left( \chi  _1 \right) = 0$, the solution for $m_0$ is the trivial choice of $m_0=0$, which also fulfills the condition $m\left( \varepsilon =0 \right) = 0$, so that for this specific case the old function given by SP08 is recovered. The solution for the higher order terms follow subsequently
\begin{align}
    m_1 &= - \int\limits _{\chi _1}^{\chi _2} b\left( \chi \right) \exp \left\lbrace - \int\limits _{\chi '}^{\chi} A\left( \chi '' \right) d\chi '' \right\rbrace d\chi '\, , \\
    m_i &= \int\limits_{\chi _1}^{\chi } \left( \sum\limits_{k=1}^{i-1} m_km_{i-k}\right) \exp \left\lbrace -\int\limits_{\chi '}^{\chi } A \left( \chi '' \right) d\chi ''  \right\rbrace d\chi ' \, .
\end{align}
In the case of diffusive escape (such as in SP08), where $\varepsilon _{\chi} = \varepsilon \chi ^{2-q}$ with a (small) constant $\varepsilon$, we can make the same calculation as above, by just simply making an expansion in $\varepsilon$ where $b\left( \chi \right) = -\chi ^{2-2q}$. \\
\\
However, it can be seen in Fig.~\ref{Fig:m_ana} that the analytical solution (colored dotted lines) starts to deviate from the correct numerical solution (black line) even for higher orders in $i$ when $A(\chi)\gtrsim 0$. So we need a different approach for energies where $A > 0$, i.e. where continuous energy losses start to dominate. If we again assume, that $B$ is a small perturbation, that fulfills $B \left( \chi \gg \chi _1 \right) \rightarrow 0$ there should exist a $\chi _{\ast}$ so that $A $ from thereon is strictly positive and $m_{\ast}$ not vanishing. the function $m\left( \chi \right)$ should then fulfill the homogeneous version of Eq.~\ref{Eq:mequation} and its solution is given by 
\begin{equation}
    m_{\rm hom}(\chi) = \frac{\exp\left( \int_{\chi_{\ast}}^\chi A(\chi')d\chi' \right)}{c_{\ast}-\int_{\chi _{\ast}}^\chi \exp\left( \int_{\chi_{\ast}}^{\chi'} A(\chi'')d\chi'' \right) d\chi'}\,,
\end{equation}
Here $c _{\ast}$ represents a constant that is determined by the value of $m\left( \chi _{\ast} \right)$, which can be determined e.g. by the method of expansion, explained above.
We notice that in this solution for $A(\chi)\gg 1$, which is typically the case for $\chi\rightarrow \chi_2$ the second term in the denominator often dominates and $m_{\rm hom}(\chi\rightarrow \chi_2)\simeq-A(\chi\rightarrow \chi_2)$. 
This can also be explained by the fact that one can rewrite the homogeneous differential equation and make a fixed-point analysis.
 For this purpose we introduce the function $\mathcal{M}_A = m/A$ and rewrite the homogeneous differential equation yielding
\begin{equation}
    \frac{d \mathcal{M} _A}{d \chi} = \mathcal{M} _A^2A\left( \chi \right) + \mathcal{M} _A A\left( \chi \right)\left( 1 - \frac{A'\left( \chi \right)}{A\left( \chi \right)} \right)\, .
\end{equation}
Hence, there is a fixed point of $\mathcal{M} _A$ e.g. under the condition that $\frac{A'}{A} \ll 1 $, which is satisfied if $A(\chi)$ is a power-law function. In that case the fixed point for $\mathcal{M} _A$ is $-1$ and the function $m$ converges against $-A$ yielding a good estimate of the high-$\chi$ behavior of $m$. Combining this insight with the expansion approximation and the analytical form of the homogeneous solution we are prepared to give a complete approximation over the whole energy range of interest (see Fig.~\ref{Fig:m_ana}) if the particle escape is not the dominating process. 
Thus, also the final ingredient of the solution (\ref{Eq:Green_final}) of the steady state momentum diffusion equation can be approximated semi-analytically. In the following we will apply our general findings to some special cases, where particles are injected mono-energetically at low and high energies, respectively.

\section{Results for some special cases}\label{Sec:Results}
\subsection{Power-laws with an arbitrary spectral index}
First, we consider straight power-law functions for $\vartheta_{\chi}=\vartheta_0\,\chi^s$ and $\varepsilon_\chi=\varepsilon_0\,\chi^{2-q}$, where $\vartheta_0, \,\varepsilon_0=\text{const}>0$ and compare our results with some of the special cases determined in SP08.
\subsubsection{Hard-sphere for various losses}
A prominent case that is due to its mathematical convenience often discussed is the so-called hard-sphere approximation, where $q=2$ and consequently $\varepsilon_\chi=\text{const}$ in case of diffusive escape. Considering relativistic electrons that also suffer from continuous synchrotron cooling an analytical solution of Eq.~\ref{Eq:BaseEquation} has already been derived by SP08. 
\begin{figure}[htb]
\centering
    \includegraphics[width=0.285\textwidth]{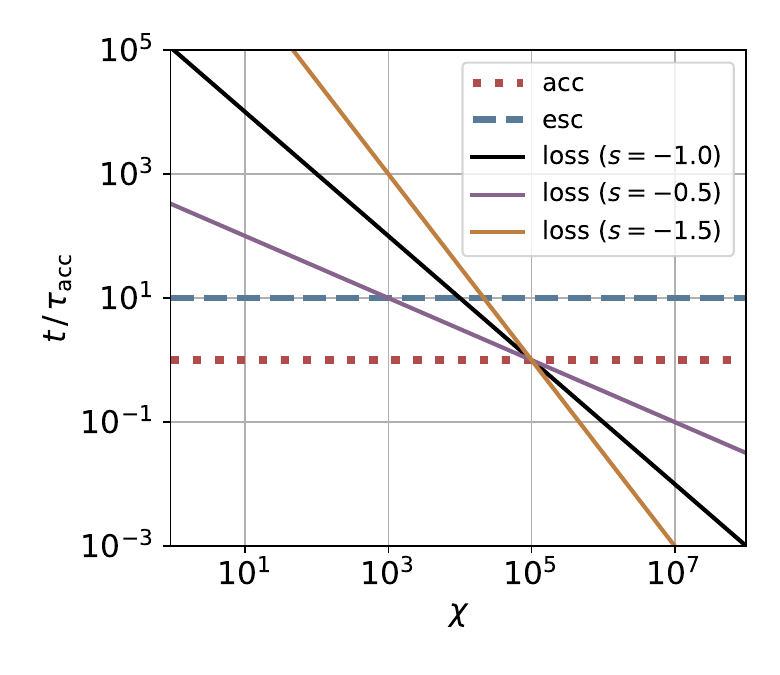}
    \includegraphics[width=0.35\textwidth]{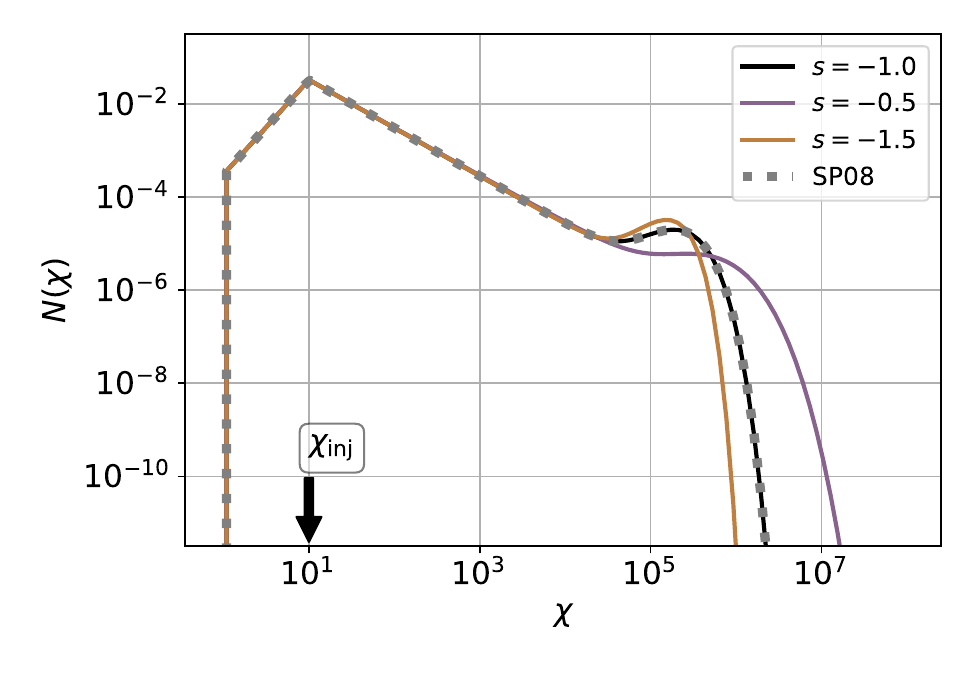}
    \includegraphics[width=0.35\textwidth]{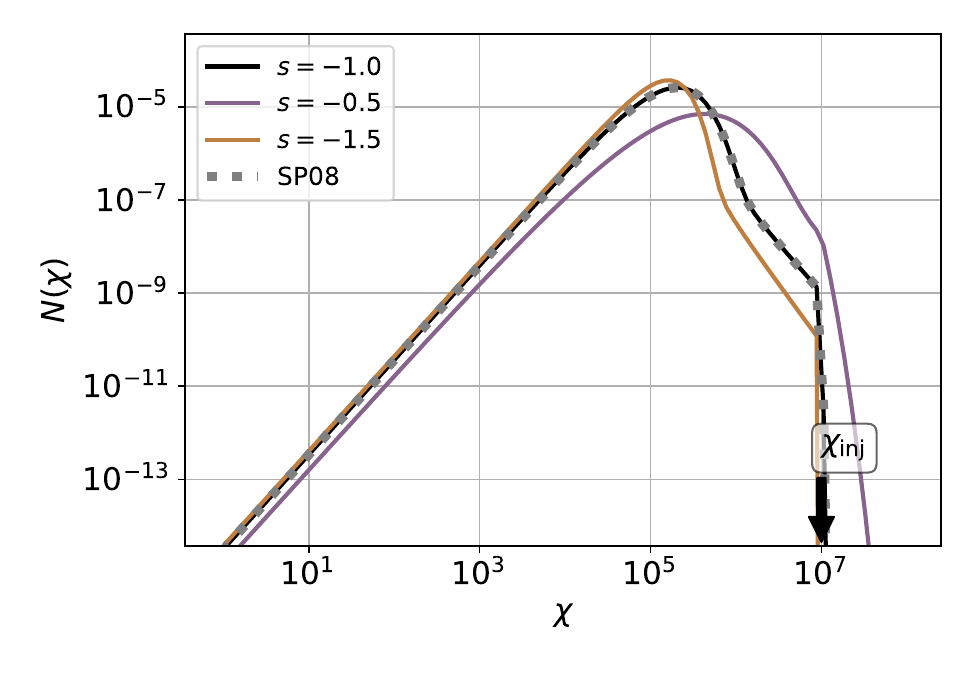}
    \includegraphics[width=0.285\textwidth]{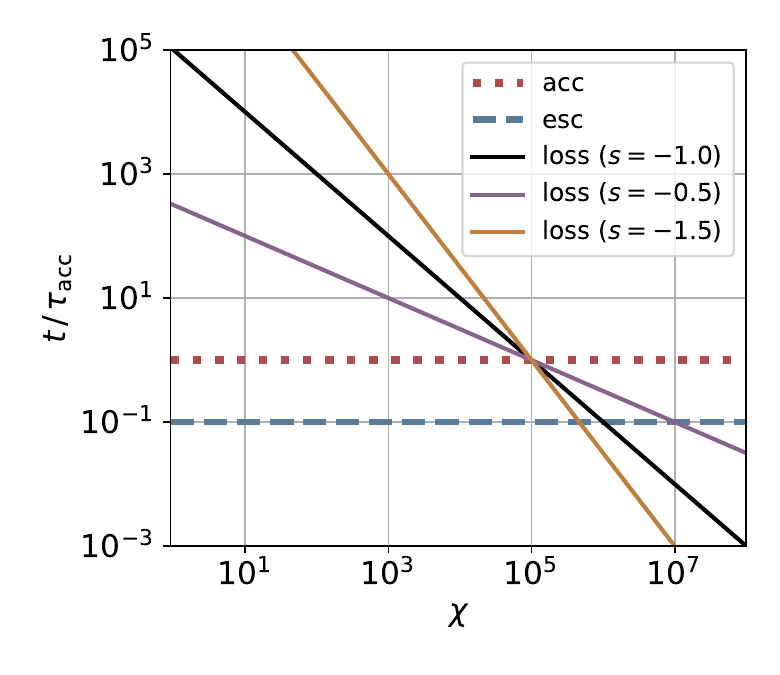}
    \includegraphics[width=0.35\textwidth]{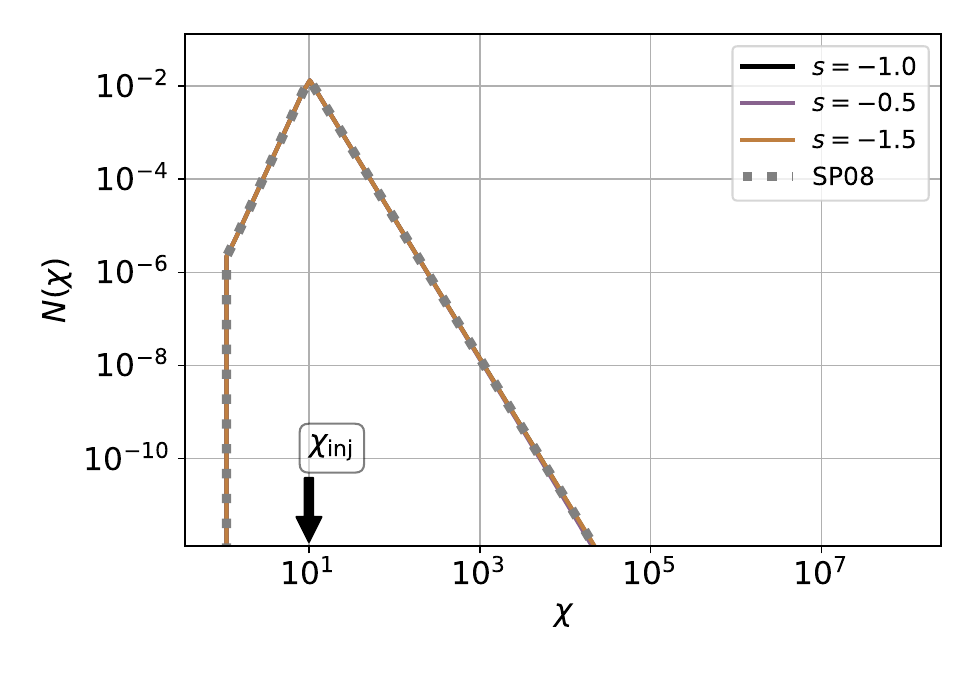}
    \includegraphics[width=0.35\textwidth]{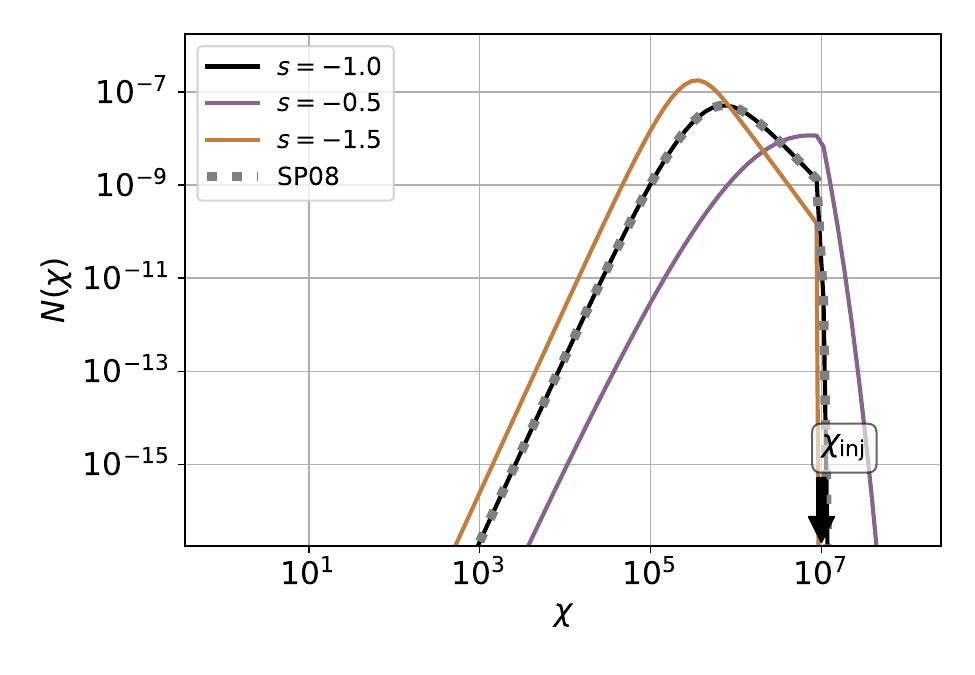}
        \caption{The characteristic timescales (left plots) and the resulting momentum distribution (middle and right plots) for the case of $q=2$ and three different unbroken power-law loss processes considering $\varepsilon_\chi=0.1$ (upper panel) and $\varepsilon_\chi=10$ (lower panel), respectively. Moreover, we used $\chi_1=0.9$, $\chi_2=5\times10^8$ and a monochromatic source rate $Q(\chi)=\delta(\chi-\chi_{\rm inj})$ with $\chi_{\rm inj}=10$ (middle plot) and $\chi_{\rm inj}=10^7$ (right plot), respectively. The continuous loss timescale is adjusted to obtain for all three cases a common intersection with the acceleration timescale. The grey dotted curves indicate the results from SP08.}
        \label{Fig:SP08-comp}
\end{figure}
The Fig.~\ref{Fig:SP08-comp} shows that we are able to exactly reproduce their previous results in case of a monochromatic injection at both low energies (middle plot) and high energies (right plot). Due to the balance of continuous cooling and acceleration at $\chi_{\rm eq}\simeq 10^5$ a pile-up bump emerges around these energies, which increases with increasing escape time. As already shown by SP08 the resulting spectral behavior can change significantly with the energy of the injected particles or more precisely we obtain two characteristic cases for $\chi_{\rm inj}\ll \chi_{\rm eq}$ and $\chi_{\rm inj}\gg \chi_{\rm eq}$, respectively. Moreover, the Fig.~\ref{Fig:SP08-comp} illustrates that the shape of this pile-up bump depends on the supposed loss processes, leading to a sharpening of the pile-up for steeper loss processes, while at energies $\chi\ll \chi_{\rm eq}$ the resulting steady state distribution does not depend on the underlying loss process for $\chi_{\rm inj}\ll \chi_{\rm eq}$. Here, a universal spectral behavior according to
\begin{equation}
    N^{q=2}_{\chi_{\rm inj}\ll \chi_{\rm eq}}(\chi)\propto \begin{cases}
        \chi^{\sigma+1}\,,\qquad &\text{for } \chi_1<\chi<\chi_{\rm inj}\,,\\
        \chi^{-\sigma}\,,\qquad &\text{for } \chi_{\rm inj}<\chi\ll \chi_{\rm eq}
    \end{cases}
\end{equation}
with $\sigma = -(1/2)+\sqrt{(9/4)+\varepsilon_{\chi}}$ is shaped as already derived by SP08. In the case of a monochromatic injection at high energies ($\chi_{\rm inj}\gg \chi_{\rm eq}$) the spectral behavior can be approximated by 
\begin{equation}
    N^{q=2}_{\chi_{\rm inj}\gg \chi_{\rm eq}}(\chi)\propto \begin{cases}
        \chi^{\sigma+1}\,,\qquad &\text{for } \chi_1<\chi\ll \chi_{\rm eq}\,,\\
        \chi^{-s-1}\,,\qquad &\text{for } \chi_{\rm eq}\ll \chi< \chi_{\rm inj}
    \end{cases}
\end{equation}
where the impact of momentum diffusion vanishes at $\chi\gg\chi_{\rm eq}$ and the high energy particles cool towards smaller energies according to the given continuous loss process. Hence, independent of the continuous loss process at work a universal spectral behavior establishes at energies $\chi\ll \chi_{\rm eq}$ that (for $q=2$) only depends on the efficiency of the particle escape from the system. 
\subsubsection{Synchrotron-like losses for various turbulence spectral indices}
\begin{figure}[htb]
\centering
    \includegraphics[width=0.285\textwidth]{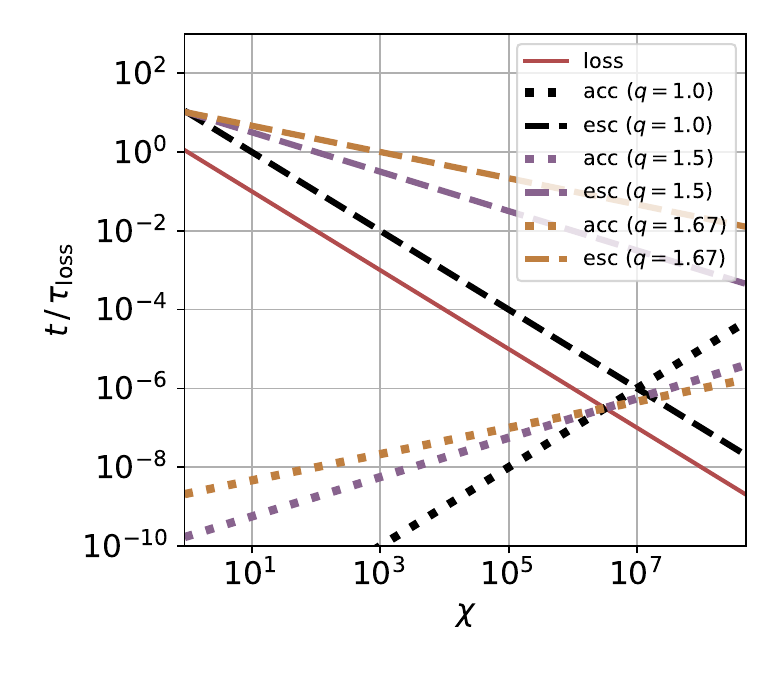}
    \includegraphics[width=0.35\textwidth]{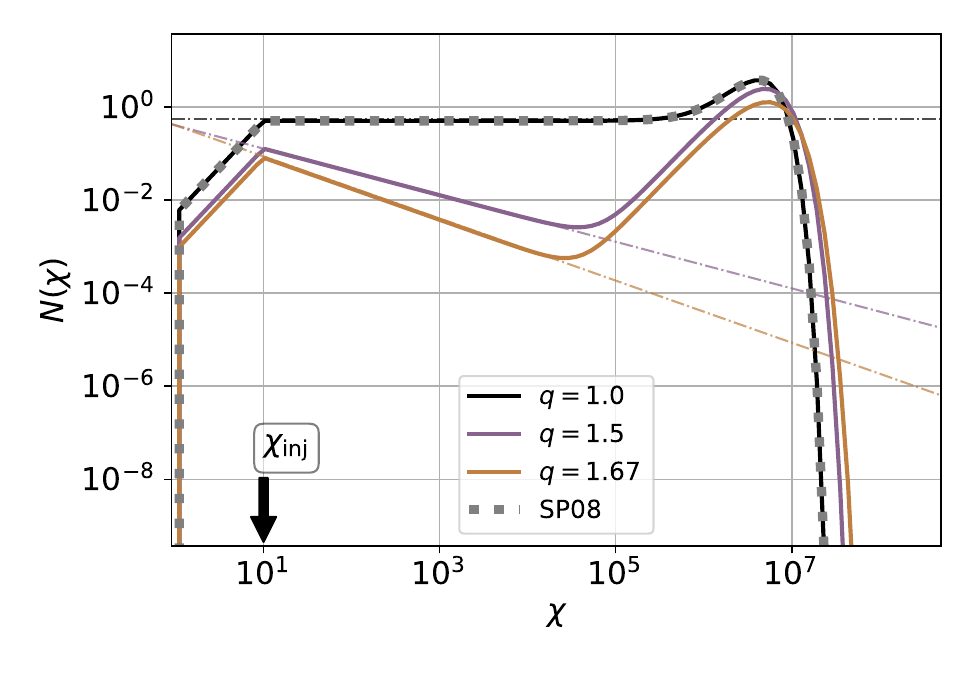}
    \includegraphics[width=0.35\textwidth]{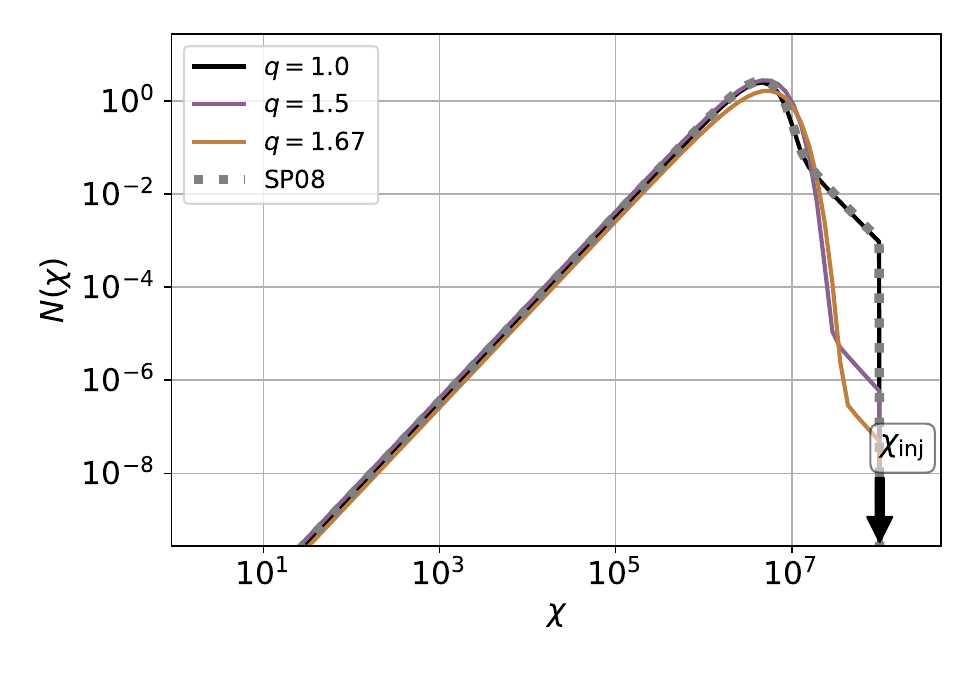}
    \includegraphics[width=0.285\textwidth]{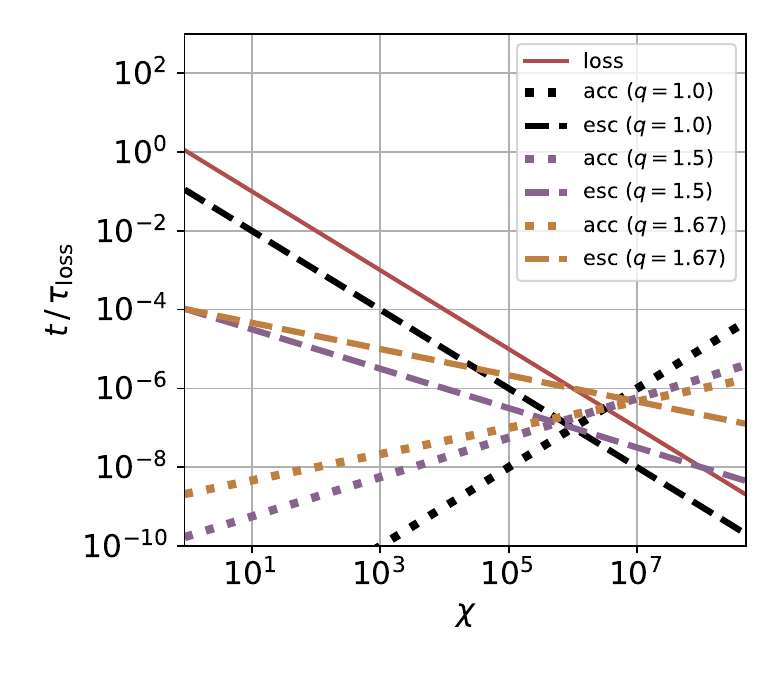}
    \includegraphics[width=0.35\textwidth]{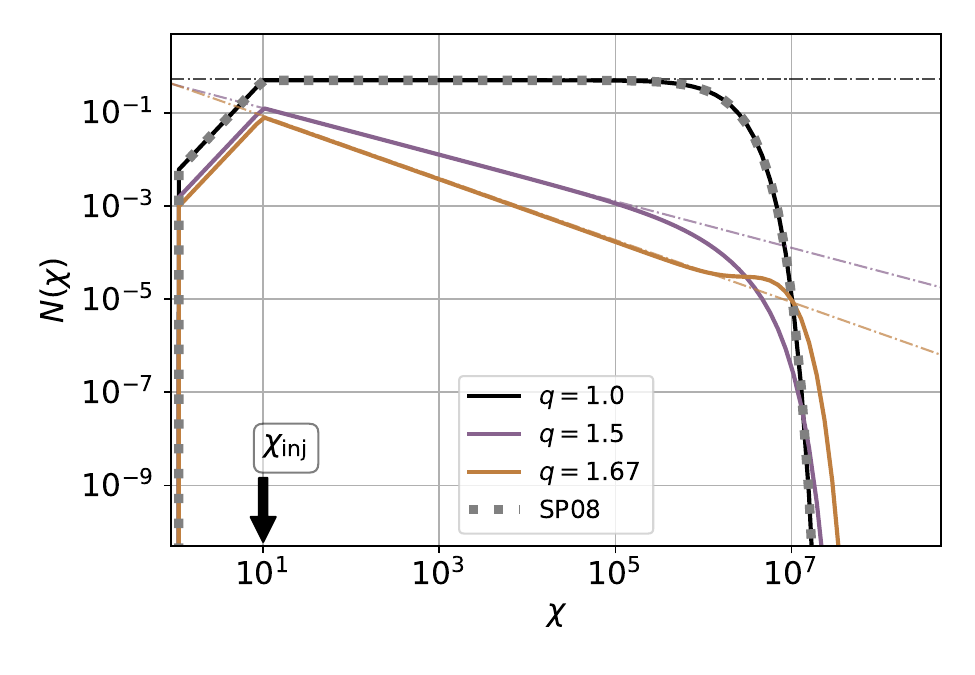}
    \includegraphics[width=0.35\textwidth]{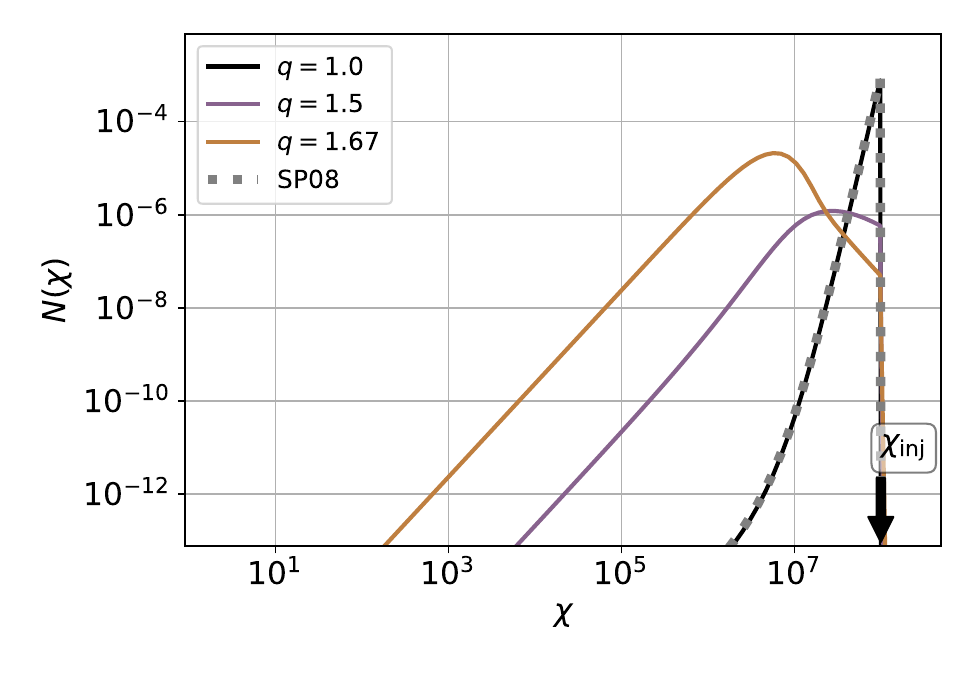}
        \caption{Same as in Fig.~\ref{Fig:SP08-comp} but for the case of $s=1$, $\varepsilon_\chi=\text{const}\ll 1$ and three different cases of the turbulence spectral index $q$ considering $\varepsilon_\chi=0.1\,\chi^{2-q}$ (upper panel) and $\varepsilon_\chi=10\,\chi^{2-q}$ (lower panel), respectively. Here, the acceleration timescale $\tau_{\rm acc}$ is adjusted to obtain for all three cases a common intersection with the continuous loss timescale. The thin dashed-dotted lines indicate the spectral behavior according to the case of pure momentum diffusion as given by $N(\chi)\propto \chi^{1-q}$.}
        \label{Fig:SP08-comp_case2}
\end{figure}

Another case that has previously already been discussed by SP08 is the Bohm limit for momentum diffusion ($q=1$) and continuous synchrotron losses ($\vartheta_\chi\propto \chi$). The Fig.~\ref{Fig:SP08-comp_case2} shows that also this case can be reproduced perfectly by our solution and further, it illustrates the difference that emerges for other spectral indices of the turbulence. We recognize that competition between energy gain and particle escape determines the spectral behavior at $\chi \ll \chi_{\rm eq}$, whereas around $\chi_{\rm eq}$ the impact of the continuous energy loss becomes relevant.  
Thus, for scenarios where $t_{\rm esc}>t_{\rm acc}$ the spectral behavior can be approximated by  
\begin{equation}
    N^{s=1}_{\chi_{\rm inj}\ll \chi_{\rm eq}}(\chi,\, t_{\rm esc}>t_{\rm acc})\propto \begin{cases}
        \chi^{2}\,,\qquad &\text{for } \chi_1<\chi<\chi_{\rm inj}\,,\\
        \chi^{1-q}\,,\qquad &\text{for } \chi_{\rm inj}<\chi\ll \chi_{\rm eq}
    \end{cases}
\end{equation}
in case of a monochromatic injection at low energies and 
\begin{equation}
    N^{s=1}_{\chi_{\rm inj}\gg \chi_{\rm eq}}(\chi,\,t_{\rm esc}>t_{\rm acc})\propto \begin{cases}
        \chi^{2}\,,\qquad &\text{for } \chi_1<\chi\ll\chi_{\rm eq}\,,\\
        \chi^{-s-1}\,,\qquad &\text{for } \chi_{\rm eq}\ll \chi< \chi_{\rm inj}
    \end{cases}
\end{equation}
for high energy particle injection. If the particle escape is faster than the acceleration process ($t_{\rm esc}<t_{\rm acc}$) the pile-up bump around $\chi_{\rm eq}$ gets suppressed and the spectral behavior at $\chi\ll \chi_{\rm eq}$ softens for low energy injection and hardens for high energy injection, respectively. Hence, the imprint of the continuous loss process vanishes if $t_{\rm esc}(\chi_{\rm eq})<t_{\rm loss}(\chi_{\rm eq})$.   
\subsection{Arbitrary first-order processes}
\begin{figure}[htb]
\centering
    \includegraphics[width=0.285\textwidth]{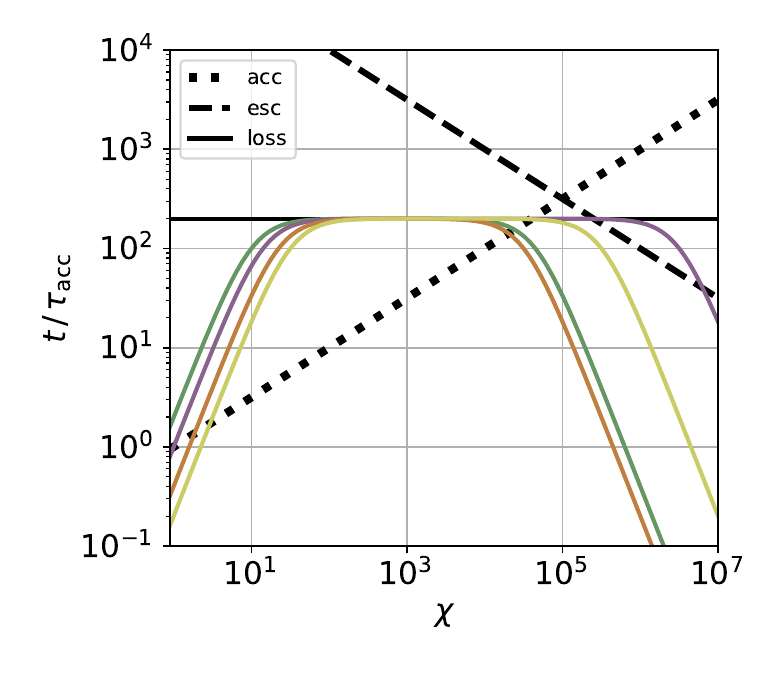}
    \includegraphics[width=0.35\textwidth]{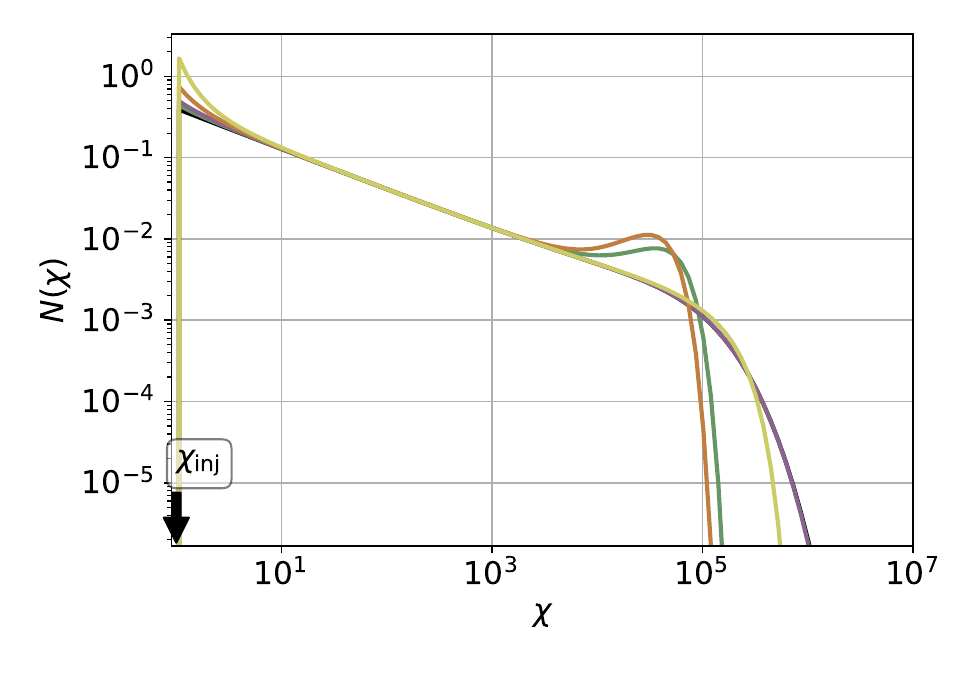}
    \includegraphics[width=0.35\textwidth]{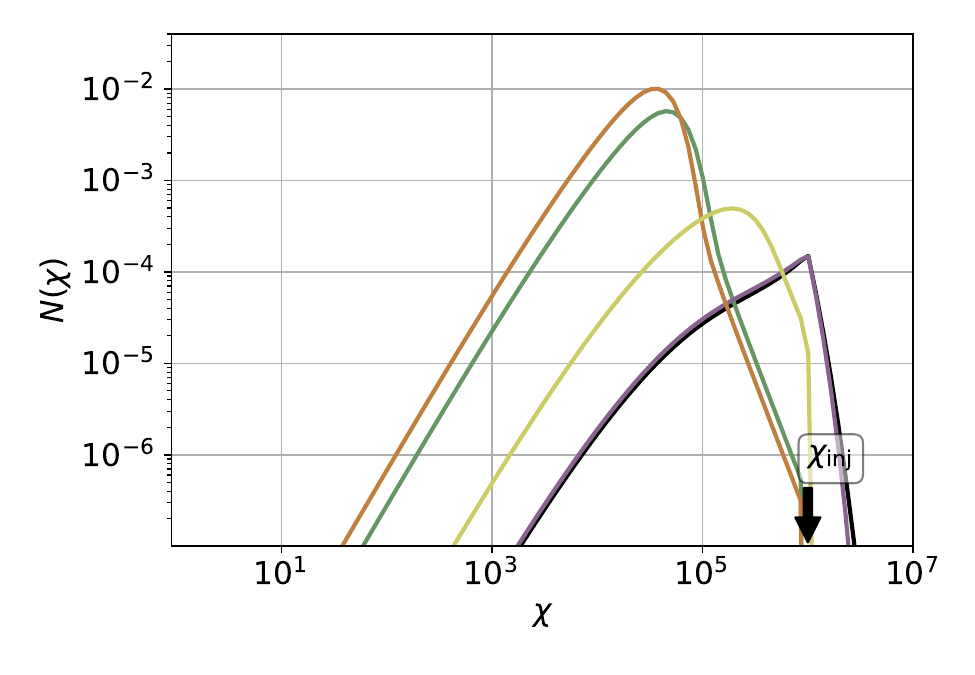}
        \caption{\emph{Left:} Characteristic timescales supposing no break (black solid line) as well as two different breaks (colored solid lines) in the continuous loss function $\vartheta_\chi$. \emph{Middle \& right:} The resulting spectral behavior in case of $q=3/2$ and a monochromatic source rate $Q(\chi)=\delta(\chi-\chi_{\rm inj})$ with $\chi_{\rm inj}=1$ (middle plot) and $\chi_{\rm inj}=10^6$ (right plot).}
        \label{Fig:ArbLoss}
\end{figure}

In most astrophysical applications the relativistic particles suffer from various loss processes, such as e.g.\ Coulomb and ionization losses that dominate at low energies and synchrotron losses dominating at high energies. Therefore $\vartheta_\chi$ is even without additional first-order energy gains rather to be described by a broken power-law. 
In the following, we first consider the exemplary case of 
\begin{equation}
    \vartheta_\chi = \tau_{\rm acc}\,\left( (\tau_{\rm loss}^{(l)})^{-1} \chi^{-2} + (\tau_{\rm loss}^{(m)})^{-1} + (\tau_{\rm loss}^{(h)})^{-1} \chi^{2}  \right) \,,
\end{equation}
which roughly mimics a combination of Coulomb plus hadronic pion production plus Bethe Heitler pair production losses as it might be realized in the AGN corona \citep[e.g.][]{Eichmann+2022}. Here, the Fig.~\ref{Fig:ArbLoss} shows similar features in the resulting energy distribution such as in the previous cases. We notice (in the middle Fig.~\ref{Fig:ArbLoss}) that the spectral behavior of $\vartheta_\chi$ at $\chi\simeq\chi_{\rm eq}$, where $\chi_{\rm eq}$ denotes the balance of acceleration and continuous loss timescales at the \emph{highest} energies, determines the characteristics of the pile-up bump at high energies. Moreover, an additional low energy pile-up bump emerges if there is another balance of these processes at lower energies. But due to the limited range of the assigned turbulence spectrum the Eq.~\ref{Eq:BaseEquation} is not appropriate at $\chi<\chi_1$ and $N(\chi<\chi_1)$ vanishes as a result of the assigned no-flux boundary condition. In the case of high-energy injection (see the right Fig.~\ref{Fig:ArbLoss}) the low-energy loss processes (at $\chi\ll \chi_{\rm eq}$) are obviously irrelevant, but the high-energy loss processes (at $\chi\gtrsim \chi_{\rm eq}$) determine the amount of particles that pile-up around $\chi_{\rm eq}$. In general the momentum distribution only depends on the spectral behavior of $\vartheta_\chi$ at the balance of acceleration and continuous losses. Though, multiple pile-up bumps can emerge when the continuous energy loss balances the acceleration at different energies. 

Finally, we consider the exemplary case of Fermi-I acceleration at low energies and two different energy loss processes at high energies so that
\begin{equation}
    \vartheta_\chi = - \tau_{\rm acc}\,\left( (\tau_{\rm gain})^{-1} \chi^{-1} - (\tau_{\rm loss}^{(m)})^{-1} - (\tau_{\rm loss}^{(h)})^{-1} \chi^{2}  \right) \,.
\end{equation}
\begin{figure}[htb]
\centering
    \includegraphics[width=0.285\textwidth]{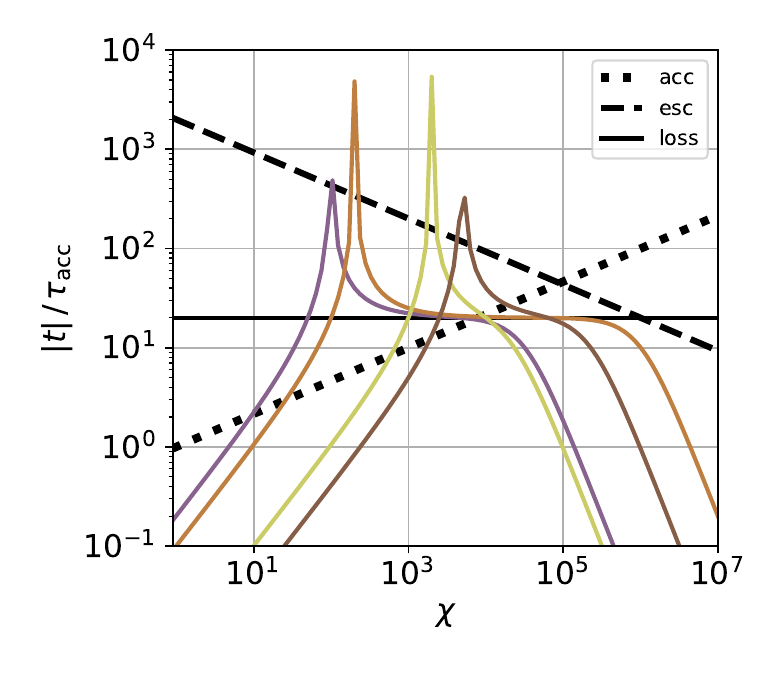}
    \includegraphics[width=0.35\textwidth]{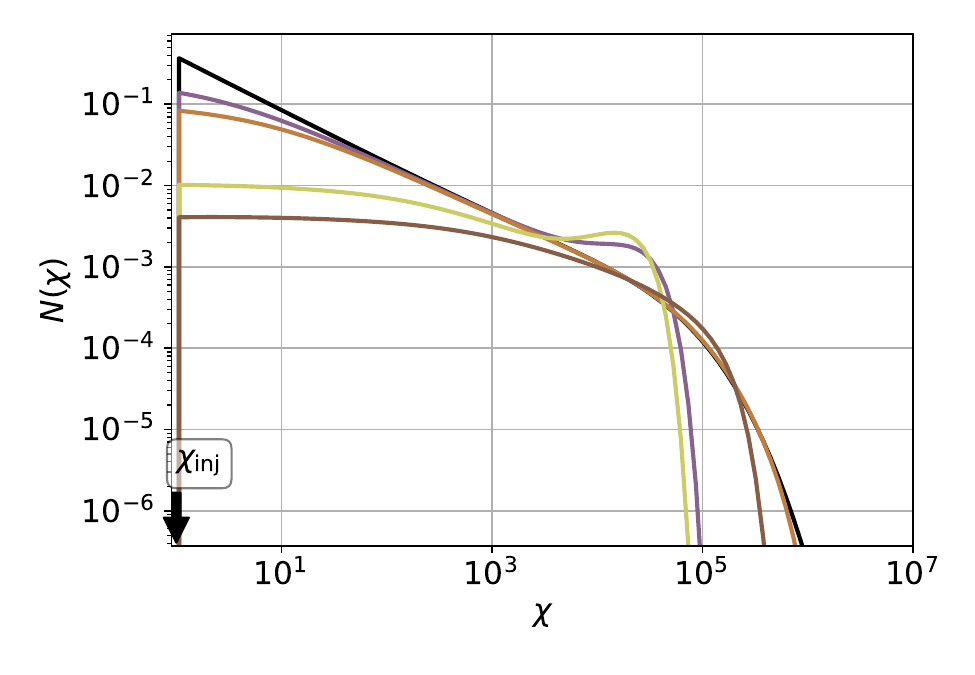}
    \includegraphics[width=0.35\textwidth]{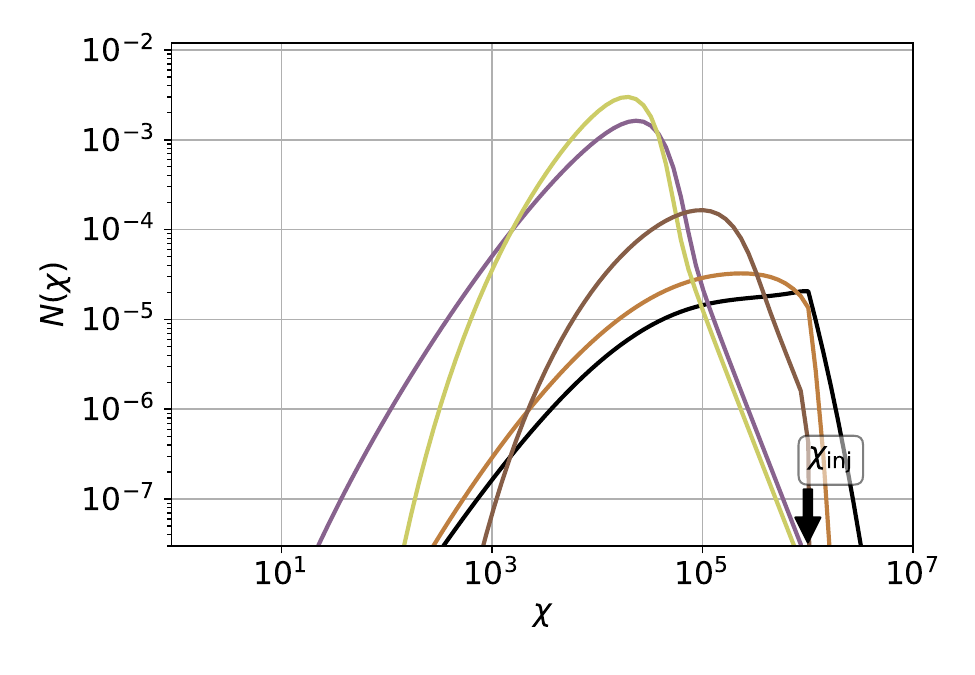}
        \caption{\emph{Left:} Absolute value of the characteristic timescales supposing no break (black solid line) as well as two different breaks (colored solid lines) in $\vartheta_\chi$, where Fermi-I processes dominate at low energies. \emph{Middle \& right:} The resulting spectral behavior in case of $q=5/3$ and a monochromatic source rate $Q(\chi)=\delta(\chi-\chi_{\rm inj})$ with $\chi_{\rm inj}=1$ (middle plot) and $\chi_{\rm inj}=10^6$ (right plot).}
        \label{Fig:ArbLossGain}
\end{figure}
Due to the additional impact by Fermi-I acceleration at low energies the middle plot in Fig.~\ref{Fig:ArbLossGain} indicates that for $\chi\lesssim \chi_{\rm tr}$, where $\chi_{\rm tr}=(\tau_{\rm gain}/\tau_{\rm acc})^{1/(1-q)}$ denotes the transition from Fermi-I to Fermi-II acceleration, the resulting spectrum becomes significantly harder reaching a constant value for $N(\chi\ll \chi_{\rm tr})$. Obviously this is not the case for an injection at energies $\chi\gg \chi_{\rm tr}$ as illustrated by the right plot in Fig.~\ref{Fig:ArbLossGain}, where particles are again piled up around $\chi_{\rm eq}$ as a result of the balance between Fermi-II acceleration and energy losses. But, still at $\chi \lesssim \chi_{\rm tr}$ Fermi-I processes yield an additional hardening of the spectrum.

\section{Summary and Outlook}\label{Sec:Summary}
Second-order Fermi acceleration is one of the key mechanism to energize CR particles. But in most astrophysical environments there are typically other competing processes, such as e.g. first-order Fermi acceleration and/or different kinds of continuous and catastrophic energy losses that become dominant at certain energies. The resulting steady state momentum distribution has however so far only been determined for a limited number of special cases that is often not adequate for the system of interest. In that case computationally intense numerical solvers had to be used to solve the problem.

We revised the SP08 approach so that the considered momentum diffusion equation (\ref{OrigTranspEq}) simplifies significantly to a first order differential equation for the so-called modification function $m(\chi)$ that treats the impact of a non-vanishing particle escape from the system. In case of effective particle escape, that however is not the leading process, we also provide an appropriate analytical approximation for $m(\chi)$. 

Using the provided Green's function (\ref{Eq:Green_final}) the steady state momentum distribution function can be determined semi-analytically for any arbitrary physical environment that features momentum diffusion in a limited range of energies as well as continuous and catastrophic energy losses or gains. Finally, we discuss the resulting momentum distribution $N(\chi)$ for some exemplary physical cases and compare our findings with the results from SP08 yielding that:
\begin{enumerate}
    \item[i)] The special cases from SP08 can be exactly reproduced. 
    \item[ii)] A pile-up bump generally emerges around the equilibrium momentum $\chi_{\rm eq}$ (of the acceleration and continuous loss processes), if the particle escape timescale $t_{\rm esc}(\chi_{\rm eq})$ is significantly larger than the acceleration/loss timescale at $\chi_{\rm eq}$. Thus, this feature is a result of the competition between stochastic energy gain and continuous energy loss. The shape of the pile-up bump agrees with previous findings for some special cases, showing a universal $\chi^2$-dependence at $\chi<\chi_{\rm eq}$ and an exponential cut-off whose steepness depends on the dominating continuous energy loss process at $\chi>\chi_{\rm eq}$.
    \item[iii)] If there are no additional first-order Fermi acceleration processes at work and a sufficient particle's residence time, we obtain $N(\chi)\propto \chi^{1-q}$ for $\chi_{\rm inj}<\chi\ll \chi_{\rm eq}$ as expected for the case of pure momentum diffusion. In case of significant first-order Fermi acceleration processes at these energies the distribution flattens for $\chi\rightarrow \chi_{\rm inj}$ to $N(\chi)\rightarrow \text{const}$. For a dominating particle escape from the system the distribution softens significantly. Hence, at $\chi_{\rm inj}<\chi\ll \chi_{\rm eq}$ the spectral behavior generally results from the competition of particle acceleration and particle escape.
    \item[iv)] If particles are injected at high energies where $\chi_{\rm inj}\gg \chi_{\rm eq}$, such as e.g.\ it could be the case for secondary electrons/positrons from hadronic interactions, the resulting momentum distribution only consists of the previously described pile-up bump with $N(\chi<\chi_{\rm eq})\propto \chi^2$ and even harder if particle escape dominates at these energies. 
\end{enumerate}
In total, we notice that the pure momentum diffusion solution with $N(\chi)\propto \chi^{1-q}$ or any other unbroken power-law function---which is often used as a simplified approximation---is only valid in a very limited set of scenarios and energies. So, to understand the resulting high-energy phenomena in calorimetric-like environments (where CR particles loose a significant fraction of their energy before leaving the system) it is indispensable to also account for the pile-up bump. Here the momentum distribution at the highest energies is typically $\propto \chi^2$ before an exponential suppression that is characterized by the dominant loss process around $\chi=\chi_{\rm eq}$ cuts it off. 

Note that in this work we only consider the case of resonant wave-particle interactions for assigned turbulence and the physical system of interest to be in steady-state. Characteristic features such as the pile-up bump can only form if the system has evolved for a sufficient time span. Hence, our findings can also be used to verify if a steady-state situations is already reached in numerical simulation with assigned turbulence \citep{MertenAerdker2024}. However, in particular at low energies---where the total energy in CR particle could exceed the assigned energy budged of the magnetic turbulence---the adopted "test particle approach" is often not appropriate anymore since the particles start damping the turbulent cascade yielding non-linear back-reactions as recently discussed by \cite{Lemoine+2024}. Moreover, it has been shown that the wave particle resonance vanishes in case of anisotropic MHD turbulence \citep{GoldreichSridhar1995, Chandran2000, YanLazarian2002}.     

However, such scenarios are beyond the scope of this work, as our approach requires assigned, isotropic turbulence in the first place. But even within these requirements there are still open questions that are not yet answered, such as the time-dependent behavior of the particle spectra or more rigid solutions of the Riccati-equation that modifies our solutions. Those are among others possible topics of future investigations.

\begin{acknowledgements}
We acknowledge funding from the German Science Foundation DFG, within
the Collaborative Research Center SFB 1491 “Cosmic Interacting Matters - From Source to Signal”. 
Moreover, we would like to thank H.~Fichtner and Y.~Litvinenko for fruitful discussions.

\end{acknowledgements}

\paragraph{Software:} 
Some of the results in this paper have been derived using the software packages Numpy \citep{vanDerWalt2011}, Scipy \citep{2020SciPy-NMeth}, Matplotlib \citep{Hunter:2007}, Seaborn \citep{michael_waskom_2017_883859}. 

\begin{appendix}

\section{General Streaming}\label{App:GenralStreaming}
In this Appendix we will go through the calculation given by Eqs. ~\ref{Eq:StreamProblem}-~\ref{Eq:StramSolution}, to derive the parameters $\alpha $ and $\beta $ for the Green's function.
We aim to determine these two parameters by assuming that in the given continuity equation (\ref{Eq:StreamProblem}) the function vanishes under the streaming operator (\ref{Op:Streaming}) at the boundaries $\chi _1$ and  $\chi _2$.

To derive the parameter values from this condition we will start by applying the operator to Eq.~\ref{Eq:GenSol}, to derive the general form of the streaming function.
In the most general form the streaming of the derived distribution function can be written as
\begin{align}
    \Cont \left( N \right) &= \Stream \left( \chi  \right) \int\limits_{\chi _1}^{\chi } \frac{u_1 \left( \chi _0 \right) u_2 \left( \chi \right)}{-w\left( \chi _0 \right) \chi _0^q} Q\left( \chi _0 \right) d\chi _0 + \Stream \left( \chi  \right)  \int\limits_{\chi }^{\chi _2}\frac{u_1 \left( \chi \right) u_2 \left( \chi _0 \right)}{-w\left( \chi _0 \right) \chi _0^q} Q\left( \chi _0 \right) d\chi _0\notag \\
    &-  \chi ^q \frac{\partial }{\partial \chi} \int\limits_{\chi _1 }^{\chi } \frac{u_1 \left( \chi _0 \right) u_2 \left( \chi \right)}{-w\left( \chi _0 \right) \chi _0^q} Q\left( \chi _0 \right) d\chi _0
    - \chi ^q \frac{\partial }{\partial \chi} \int\limits_{\chi _1 }^{\chi } \frac{u_1 \left( \chi  \right) u_2 \left( \chi _0\right)}{-w\left( \chi _0 \right) \chi _0^q} Q\left( \chi _0 \right) d\chi _0 \\
    &= \Stream \left( \chi  \right)  u_2 \left( \chi \right)\int\limits_{\chi _1}^{\chi } \frac{u_1 \left( \chi _0 \right)}{-w\left( \chi _0 \right) \chi _0^q} Q\left( \chi _0 \right) d\chi _0 + \Stream \left( \chi  \right)  u_1 \left( \chi \right)\int\limits_{\chi }^{\chi _2}\frac{ u_2 \left( \chi _0 \right)}{-w\left( \chi _0 \right) \chi _0^q} Q\left( \chi _0 \right) d\chi _0 \notag \\
    &- \left( \chi ^q \frac{\partial u_2 \left( \chi  \right) }{\partial \chi} \right) \int\limits_{\chi_ 1}^{\chi } \frac{u_1 \left( \chi _0 \right)}{-w\left( \chi _0 \right) \chi _0^q} Q\left( \chi _0 \right) d\chi _0 
    - \left( \chi ^q \frac{\partial u_1 \left( \chi  \right)}{\partial \chi}  \right)\int\limits_{\chi }^{\chi _2} \frac{ u_2 \left( \chi _0 \right)}{-w\left( \chi _0 \right) \chi _0^q} Q\left( \chi _0 \right) d\chi _0 \\
    &- \chi ^q u_2 \left( \chi  \right) \frac{u_1 \left( \chi \right)}{- w\left( \chi \right) \chi ^q} Q \left( \chi \right) + \chi ^q u_1 \left( \chi  \right) \frac{u_2 \left( \chi \right)}{- w\left( \chi \right) \chi ^q} Q \left( \chi \right)\\
    &= \Cont \left( u_2 \right) \int\limits_{\chi _1}^{\chi } \frac{u_1 \left( \chi _0 \right)}{-w\left( \chi _0 \right) \chi _0^q} Q\left( \chi _0 \right) d\chi _0 + \Cont \left( u_1 \right) \int\limits_{\chi }^{\chi _2} \frac{ u_2 \left( \chi _0 \right)}{-w\left( \chi _0 \right) \chi _0^q} Q\left( \chi _0 \right) d\chi _0
\end{align}
and, therefore, we obtain at the boundaries, that
\begin{align}
    \Cont \left( N \right ) \mid _{\chi _1} &= \Cont \left( u_1 \right) \mid _{\chi _1} \int\limits_{\chi _1 }^{\chi _2} \frac{ u_2 \left( \chi _0 \right)}{-w\left( \chi _0 \right) \chi _0^q} Q\left( \chi _0 \right) d\chi _0 \label{Eq:LeftFlow}\\
    \Cont \left( N \right ) \mid _{\chi _2} &= \Cont \left( u_2 \right) \mid _{\chi _2} \int\limits_{\chi _1}^{\chi _2} \frac{u_1 \left( \chi _0 \right)}
    {-w\left( \chi _0 \right) \chi _0^q} Q\left( \chi _0 \right) d\chi _0 \, .\label{Eq:RightFlow}
\end{align}
Since $u_1$ and $u_2$ are identical when swapping $\alpha$ for $\beta$, we can subsequently introduce the general case of $u = y_1 + \gamma y_2$, where $\gamma\in\{\alpha,\,\beta\}$ and
\begin{align}
    y_1 &= \tilde{S}^{-1}\,, \\
    y_2 &= y_1 \int\limits_{\chi _1}^{ \chi } \chi _0^{-q}\tilde{S } \exp \left\lbrace \int\limits_{\chi _1}^{\chi _0} m \left( \chi _{00} \right) d\chi_{00} \right\rbrace d\chi _0\, , \\
    \tilde{S} &= S \exp \left\lbrace \int\limits_{\chi _1}^{\chi } m \left( \chi _{0} \right)d\chi _0 \right\rbrace\, , \\
    S &= \exp \left\lbrace \int\limits_{\chi _1 }^{ \chi } -2\chi _0^{-1}+ \chi _0^{1-q} \theta _{\chi} d\chi _0 \right\rbrace \, . 
\end{align}

Using
\begin{align}
    \Cont \left( y_1 \right) &=  \chi ^q  \tilde{S}^{-2}  \left(   \exp \left\lbrace \int\limits_{\chi _1}^{\chi } m \left( \chi _0 \right) d\chi _0 \right\rbrace \frac{\partial S}{\partial  \chi } + m \left( \chi  \right)\tilde{S} \right)+ \left(  2 \chi ^{q-1} - \chi \theta _{\chi }\right) \tilde{S}^{-1} \notag \\
     &=  \chi ^q \tilde{S}^{-2}  \left(  - \exp \left\lbrace \int\limits_{\chi _1}^{\chi } m \left( \chi _0 \right) d\chi _0 \right\rbrace \left( 2\chi ^{-1} - \chi ^{1-q}\theta _{\chi} \right) S  + m \left( \chi \right) \tilde{S} \right)+ \left(  2 \chi ^{q-1} - \chi \theta _{\chi }\right) \tilde{S}^{-1} \notag \\
     &=   \left( -  \left( 2\chi ^{q-1} - \chi \theta _{\chi} \right) \tilde{S}^{-1}  + m \left( \chi  \right) \chi ^q \tilde{S}^{-1} \right)+ \left(  2 \chi ^{q-1} - \chi \theta _{\chi }\right) \tilde{S}^{-1} \notag \\
     &= m \left( \chi  \right) \chi ^q y_1 \, ,
\end{align}
as well as
\begin{align}
    \Cont \left( y_2 \right) &= \Cont \left( y_1 \right) \int\limits_{\chi _1}^{ \chi } \chi _0^{-q}\tilde{S } \exp \left\lbrace \int\limits_{\chi _1}^{\chi _0} m \left( \chi _{00} \right)d\chi_{00} \right\rbrace d\chi _0  - y_1 \chi ^q \chi ^{-q} \tilde{S} \exp \left\lbrace \int\limits_{\chi _1}^{\chi } m \left( \chi _{0} \right) d\chi_{0} \right\rbrace \\
    &= m  \left( \chi  \right)\chi ^q y_1 \int\limits_{\chi _1}^{ \chi } \chi _0^{-q}\tilde{S } \exp \left\lbrace \int\limits_{\chi _1}^{\chi _0} m \left( \chi _{00} \right) d\chi_{00} \right\rbrace d\chi _0
    -  \exp \left\lbrace \int\limits_{\chi _1}^{\chi } m \left( \chi _{0} \right) d\chi_{0} \right\rbrace \\
    &= m \left( \chi  \right) \chi ^q y_2 
    -  \exp \left\lbrace \int\limits_{\chi _1}^{\chi } m \left( \chi _{0} \right) d\chi_{0} \right\rbrace \, ,
\end{align}
we can subsequently determine, that
\begin{align}
    \Cont \left( u \right) &= \Cont \left( y_1 \right) + \gamma \Cont \left( y_2 \right) \\
    &= m \left( \chi  \right) \chi ^q y_1 + \gamma m \left( \chi  \right) \chi ^q y_2  - \gamma   \exp \left\lbrace \int\limits_{\chi _1}^{\chi } m \left( \chi _{0} \right) d\chi_{0} \right\rbrace  \\
    &= m \left( \chi  \right) \chi ^q u -  \gamma   \exp \left\lbrace \int\limits_{\chi _1}^{\chi } m \left( \chi _{0} \right) d\chi_{0} \right\rbrace \, .
\end{align}
Hence, for the special cases we are considering follows that
\begin{align}
    \Cont \left( u_1 \right) &= m \left( \chi \right) \chi ^q u_1 \left( \chi \right) - \alpha \exp \left\lbrace \int\limits_{\chi _1}^{\chi} m \left( \chi _0 \right) d\chi _0 \right\rbrace \label{Eq:StreamU1}\\
    \Cont \left( u_2 \right) &=   m \left( \chi \right)\chi ^q u_2 \left( \chi \right) - \beta \exp \left\lbrace \int\limits_{\chi _1}^{\chi} m \left( \chi _0 \right) d\chi _0 \right\rbrace \, .\label{Eq:StreamU2}
\end{align}
The two free parameters $\alpha$ and $\beta$ now determine the fluxes at $\chi _1$ and $\chi _2$. We choose the option of vanishing fluxes at both boundaries, so the injected particles by the source are completely outdone by the loss term that depends on $\varepsilon$.
So we set Eq.(\ref{Eq:LeftFlow}) and Eq.(\ref{Eq:RightFlow}) to zero and determine the necessary values of $\alpha$ and $\beta$ from there. For the sake of simplicity we start with Eq.(\ref{Eq:LeftFlow}) and insert Eq.(\ref{Eq:StreamU1}) at $\chi _1$.
It follows:
\begin{align}
    0 \overset{!}{=}  \frac{\alpha}{\beta - \alpha }  \left( \int\limits_{\chi_1}^{\chi_2} y_1\left( \chi_0 \right) S \left( \chi_0 \right) d\chi_0 + \beta \int\limits_{\chi_1}^{\chi_2} y_1\left( \chi_0 \right) S \left( \chi_0 \right) d\chi_0 \right) \, .\label{Eq:AlpBetDet}
\end{align}
There we used the  Wronskian, which  can be calculated by the following means:
\begin{align}
    w &= u_1u_2' - u_1'u_2 \\
    &= \left( y_1 + \alpha y_2 \right) \left( y_1' + \beta y_2' \right) - \left( y_1' + \alpha y_2' \right) \left( y_1 + \beta y_2 \right) \\
    &= \left( \beta - \alpha \right) \left( y_1y_2' - y_1'y_2 \right) \\
    &= \left( \beta - \alpha \right) \left( y_1 y_1' \int\limits_{\chi _1}^{\chi } \ldots d\chi _0 + y_1^2 \chi ^{-q} \tilde{S} \exp \left\lbrace \int\limits_{\chi _1}^{\chi } \ldots d\chi _0 \right\rbrace y_1 y_1' \int\limits_{\chi _1}^{\chi } \ldots d\chi _0 \right) \\
    &= \left( \beta - \alpha \right) \left( \tilde{S}^{-1} \chi ^{-q} \exp \left\lbrace \int\limits_{\chi _1 }^{\chi } \ldots d\chi_0 \right\rbrace \right) \\
    &= \left( \beta - \alpha \right) \left( S^{-1} \chi ^{-q} \right) \, . \\
\end{align}
There are two possible solutions for Eq.(\ref{Eq:AlpBetDet}), the trivial solution 
$\alpha = 0$ and the solution 
\begin{equation}
   \beta = - \frac{ \int\limits_{\chi_1}^{\chi_2} y_1\left( \chi_0 \right) S \left( \chi_0 \right) d\chi_0}{ \int\limits_{\chi _1}^{\chi_2} y_2\left( \chi_0 \right) S \left( \chi_0 \right) d\chi_0 } . 
\end{equation}
We will now insert both of these two possible solutions into Eq.(\ref{Eq:RightFlow}), so that we can see, which of these solutions will provide us with a consistent solution for the other parameter. If we take the solution obtained for $\beta$ and insert it into Eq.(\ref{Eq:RightFlow})) with the addition of Eq.(\ref{Eq:StreamU2}) at $\chi _2$ it follows that:
\begin{equation}
    \left( m \left( \chi _2 \right) \chi _2^q u_2\left( \chi _2 \right)- \beta \exp \left\lbrace \int\limits_{\chi _1}^{\chi _2} m \left( \chi _0 \right) d\chi _0 \right\rbrace \right)  \int\limits _{\chi _1}^{\chi _2} y_1\left( \chi _0 \right) S \left( \chi _0 \right) d\chi _0 = 0\, .
\end{equation}
Because the first bracket does not necessarily vanish with the given value for $\beta$, we must assume, that again the integral vanishes. This, however, will directly lead to the solution
\begin{equation}
    \alpha = -\frac{\int\limits_{\chi_1}^{\chi_2} y_1\left( \chi_0 \right) S \left( \chi_0 \right) d\chi_0}{\int\limits_{\chi_1}^{\chi_2} y_2\left( \chi_0 \right) S \left( \chi_0 \right) d\chi_0} = \beta\, .
\end{equation}
This would not only let the factor $(\beta - \alpha)^{-1}$ diverge, it would also indicate that our solutions $u_1$ and $u_2$ are the same solutions, which goes directly against our assumption going into the whole calculation.
We must, therefore, turn our attention to the solution $\alpha =0$. Inserting this into Eq.(\ref{Eq:RightFlow}) we end up with the condition, that 
\begin{equation}
    m \left( \chi _2 \right) \chi _2 ^q y_1 \left( \chi _2 \right) + \beta m \left( \chi _2 \right)\chi _2 ^q  y_2 \left( \chi _2 \right) + \beta \exp \left\lbrace \int\limits _{\chi _1}^{\chi _2} m \left( \chi _0 \right) d\chi _0 \right\rbrace \overset{!}{=}0 \, .
\end{equation}
This leads straightforwardly to
\begin{equation}
    \beta = \frac{m \left( \chi _2 \right)\chi _2 ^q  y_1 \left( \chi _2 \right)}{\exp \left\lbrace \int\limits _{\chi _1}^{\chi _2} m \left( \chi _0 \right) d\chi _0 \right\rbrace - m \left( \chi _2 \right) \chi _2 ^q  y_2 \left( \chi _2 \right)}\,.
\end{equation}
 Remember, that we have chosen $m \left( \chi _1 \right) $ to be zero. But we could have chosen any arbitrary value, which would however, change the values of $\alpha$ and $\beta$, leading to the same Green's function in the end. 

\end{appendix}

\bibliography{Bib}{}

\begin{thebibliography}{34}
\expandafter\ifx\csname natexlab\endcsname\relax\def\natexlab#1{#1}\fi

\bibitem[{{Achterberg}(1979)}]{Achterberg1979}
{Achterberg}, A. 1979, \aap, 76, 276

\bibitem[{{Chandran}(2000)}]{Chandran2000}
{Chandran}, B. D.~G. 2000, \prl, 85, 4656

\bibitem[{Comisso \& Sironi(2018)}]{ComissoSironi2018}
Comisso, L. \& Sironi, L. 2018, Phys. Rev. Lett., 121, 255101

\bibitem[{{Davis}(1956)}]{Davis1956}
{Davis}, L. 1956, Physical Review, 101, 351

\bibitem[{{Droege} \& {Schlickeiser}(1986)}]{DroegeSchlickeiser1986}
{Droege}, W. \& {Schlickeiser}, R. 1986, \apj, 305, 909

\bibitem[{Drury(1983)}]{Drury1983}
Drury, L.~O. 1983, Rept. Prog. Phys., 46, 973

\bibitem[{Eichmann {et~al.}(2022)Eichmann, Oikonomou, Salvatore, Dettmar, \&
  Tjus}]{Eichmann+2022}
Eichmann, B., Oikonomou, F., Salvatore, S., Dettmar, R.-J., \& Tjus, J.~B.
  2022, The Astrophysical Journal, 939, 43

\bibitem[{{Ellison} {et~al.}(1990){Ellison}, {Reynolds}, \&
  {Jones}}]{Ellison1990}
{Ellison}, D.~C., {Reynolds}, S.~P., \& {Jones}, F.~C. 1990, \apj, 360, 702

\bibitem[{{Fermi}(1949)}]{Fermi1949}
{Fermi}, E. 1949, Physical Review, 75, 1169

\bibitem[{{Fiorillo} {et~al.}(2024){Fiorillo}, {Comisso}, {Peretti},
  {Petropoulou}, \& {Sironi}}]{Fiorillo+2024}
{Fiorillo}, D. F.~G., {Comisso}, L., {Peretti}, E., {Petropoulou}, M., \&
  {Sironi}, L. 2024, arXiv e-prints, arXiv:2407.01678

\bibitem[{{Goldreich} \& {Sridhar}(1995)}]{GoldreichSridhar1995}
{Goldreich}, P. \& {Sridhar}, S. 1995, \apj, 438, 763

\bibitem[{Hunter(2007)}]{Hunter:2007}
Hunter, J.~D. 2007, Comput. Sci. Eng., 9, 90

\bibitem[{{IceCube Collaboration}(2022)}]{IceCube2022_ngc1068}
{IceCube Collaboration}. 2022, Science, 378, 538

\bibitem[{{Katarzy{\'n}ski} {et~al.}(2006){Katarzy{\'n}ski}, {Ghisellini},
  {Tavecchio}, {Gracia}, \& {Maraschi}}]{Kataruski}
{Katarzy{\'n}ski}, K., {Ghisellini}, G., {Tavecchio}, F., {Gracia}, J., \&
  {Maraschi}, L. 2006, \mnras, 368, L52

\bibitem[{Kolmogorov(1941)}]{Kolmogorov1941}
Kolmogorov, A.~N. 1941, Doklady Akademii Nauk SSSR, 30, 301, reprinted in:
  Proceedings of the Royal Society A, 1991, 434:9-13

\bibitem[{Kraichnan(1965)}]{Kraichnan1965}
Kraichnan, R.~H. 1965, Physics of Fluids, 8, 1385

\bibitem[{{Lacombe}(1979)}]{Lacombe1979}
{Lacombe}, C. 1979, \aap, 71, 169

\bibitem[{{Lazarian} \& {Vishniac}(1999)}]{LazarianVishniac1999}
{Lazarian}, A. \& {Vishniac}, E.~T. 1999, \apj, 517, 700

\bibitem[{{Lemoine} {et~al.}(2024){Lemoine}, {Murase}, \&
  {Rieger}}]{Lemoine+2024}
{Lemoine}, M., {Murase}, K., \& {Rieger}, F. 2024, \prd, 109, 063006

\bibitem[{{Liu} {et~al.}(2006){Liu}, {Petrosian}, {Melia}, \&
  {Fryer}}]{Liu2006}
{Liu}, S., {Petrosian}, V., {Melia}, F., \& {Fryer}, C.~L. 2006, \apj, 648,
  1020

\bibitem[{{Melrose}(1968)}]{Melrose1968}
{Melrose}, D.~B. 1968, \apss, 2, 171

\bibitem[{{Merten} \& {Aerdker}(2024)}]{MertenAerdker2024}
{Merten}, L. \& {Aerdker}, S. 2024, arXiv e-prints, arXiv:2410.01472

\bibitem[{{Murase} {et~al.}(2020){Murase}, {Kimura}, \&
  {M{\'e}sz{\'a}ros}}]{Murase+2020}
{Murase}, K., {Kimura}, S.~S., \& {M{\'e}sz{\'a}ros}, P. 2020, \prl, 125,
  011101

\bibitem[{{Park} \& {Petrosian}(1995)}]{ParkPetrosian1995}
{Park}, B.~T. \& {Petrosian}, V. 1995, \apj, 446, 699

\bibitem[{{Petrosian} \& {Donaghy}(1999)}]{Petrosian1999}
{Petrosian}, V. \& {Donaghy}, T.~Q. 1999, \apj, 527, 945

\bibitem[{{Pezzi} {et~al.}(2022){Pezzi}, {Blasi}, \& {Matthaeus}}]{Pezzi+2022}
{Pezzi}, O., {Blasi}, P., \& {Matthaeus}, W.~H. 2022, \apj, 928, 25

\bibitem[{{Schlickeiser}(1989)}]{Schlickeiser1989}
{Schlickeiser}, R. 1989, \apj, 336, 243

\bibitem[{{Stawarz} \& {Petrosian}(2008)}]{StawPetro2008}
{Stawarz}, {\L}. \& {Petrosian}, V. 2008, \apj, 681, 1725

\bibitem[{{Trotta} {et~al.}(2020){Trotta}, {Franci}, {Burgess}, \&
  {Hellinger}}]{Trotta+2020}
{Trotta}, D., {Franci}, L., {Burgess}, D., \& {Hellinger}, P. 2020, \apj, 894,
  136

\bibitem[{van~der Walt {et~al.}(2011)van~der Walt, Colbert, \&
  Varoquaux}]{vanDerWalt2011}
van~der Walt, S., Colbert, S.~C., \& Varoquaux, G. 2011
  [\eprint[arXiv]{1102.1523}]

\bibitem[{Virtanen {et~al.}(2020)Virtanen, Gommers, Oliphant, Haberland, Reddy,
  Cournapeau, Burovski, Peterson, Weckesser, Bright, {van der Walt}, Brett,
  Wilson, Millman, Mayorov, Nelson, Jones, Kern, Larson, Carey, Polat, Feng,
  Moore, {VanderPlas}, Laxalde, Perktold, Cimrman, Henriksen, Quintero, Harris,
  Archibald, Ribeiro, Pedregosa, {van Mulbregt}, \& {SciPy 1.0
  Contributors}}]{2020SciPy-NMeth}
Virtanen, P., Gommers, R., Oliphant, T.~E., {et~al.} 2020, Nature Methods, 17,
  261

\bibitem[{Waskom {et~al.}(2017)Waskom, Botvinnik, O'Kane, Hobson, Lukauskas,
  Gemperline, Augspurger, Halchenko, Cole, Warmenhoven, de~Ruiter, Pye, Hoyer,
  Vanderplas, Villalba, Kunter, Quintero, Bachant, Martin, Meyer, Miles, Ram,
  Yarkoni, Williams, Evans, Fitzgerald, Brian, Fonnesbeck, Lee, \&
  Qalieh}]{michael_waskom_2017_883859}
Waskom, M., Botvinnik, O., O'Kane, D., {et~al.} 2017, {mwaskom/seaborn: v0.8.1
  (September 2017)}

\bibitem[{{Xu} \& {Lazarian}(2023)}]{XuLazarian2023}
{Xu}, S. \& {Lazarian}, A. 2023, \apj, 942, 21

\bibitem[{{Yan} \& {Lazarian}(2002)}]{YanLazarian2002}
{Yan}, H. \& {Lazarian}, A. 2002, \prl, 89, 281102

\end{thebibliography}
\bibliographystyle{aa}
\end{document}